\documentclass[sigconf,9pt]{acmart}

\usepackage{amsmath,amsfonts}
\usepackage{algorithmic}
\usepackage{graphicx}
\usepackage{textcomp}
\usepackage{xcolor}
\usepackage{multirow}
\usepackage{gensymb}
\usepackage{subcaption}

\AtBeginDocument{%
  }

\setcopyright{acmlicensed}

\acmYear{2024}\copyrightyear{2024}
\acmConference[SenSys '24]{ACM Conference on Embedded Networked Sensor Systems}{November 4--7, 2024}{Hangzhou, China}
\acmBooktitle{ACM Conference on Embedded Networked Sensor Systems (SenSys '24), November 4--7, 2024, Hangzhou, China}
\acmDOI{10.1145/3666025.3699323}
\acmISBN{979-8-4007-0697-4/24/11}

\begin{document}

\title{SALINA: Towards Sustainable Live Sonar Analytics \\ in Wild Ecosystems}

\settopmatter{authorsperrow=4}
\author{Chi Xu}
\affiliation{%
  \institution{Simon Fraser University}
  \city{Burnaby}
  \country{Canada}}
\email{chix@sfu.ca}

\author{Rongsheng Qian}
\affiliation{%
  \institution{Simon Fraser University}
    \city{Burnaby}
  \country{Canada}}
\email{rqa4@sfu.ca}

\author{Hao Fang}
\affiliation{%
  \institution{Simon Fraser University}
    \city{Burnaby}
  \country{Canada}}
\email{fanghaof@sfu.ca}

\author{Xiaoqiang Ma}
\affiliation{%
  \institution{Douglas College}
    \city{New Westminster}
  \country{Canada}}
\email{mxqcs@ieee.org}

\author{William I. Atlas}
\affiliation{%
  \institution{Wild Salmon Center}
  \city{Portland}
  \country{United States}}
\email{watlas@wildsalmoncenter.org}

\author{Jiangchuan Liu}
\affiliation{%
  \institution{Simon Fraser University}
    \city{Burnaby}
  \country{Canada}}
\email{jcliu@sfu.ca}

\author{Mark A. Spoljaric}
\affiliation{%
  \institution{Haida Fisheries Program}
    \city{Skidegate}
  \country{Canada}}
\email{mark.spoljaric@haidanation.com}


\begin{abstract}


Sonar radar captures visual representations of underwater objects and structures using sound wave reflections, making it essential for exploration, mapping, and continuous surveillance in wild ecosystems. Real-time analysis of sonar data is crucial for time-sensitive applications, including environmental anomaly detection and in-season fishery management, where rapid decision-making is needed. However, the lack of both relevant datasets and pre-trained DNN models, coupled with resource limitations in wild environments, hinders the effective deployment and continuous operation of live sonar analytics.

We present SALINA, a sustainable live sonar analytics system designed to address these challenges. SALINA enables real-time processing of acoustic sonar data with spatial and temporal adaptations, and features energy-efficient operation through a robust energy management module. Deployed for six months at two inland rivers in British Columbia, Canada, SALINA provided continuous 24/7 underwater monitoring, supporting fishery stewardship and wildlife restoration efforts. Through extensive real-world testing, SALINA demonstrated an up to 9.5\% improvement in average precision and a 10.1\% increase in tracking metrics. The energy management module successfully handled extreme weather, preventing outages and reducing contingency costs. These results offer valuable insights for long-term deployment of acoustic data systems in the wild.


\end{abstract}









\maketitle

\section{Introduction}

Sonar radar employs sound wave reflections to visualize objects and structures within its detection range. This technology proves invaluable across a wide range of sensing applications, including underwater exploration and mapping, continuous surveillance, and studying marine life~\cite{bongiovanni2022high, kay2022caltech}. Typical analytical tasks include object detection~\cite{Steiniger2022Survey}, counting~\cite{liu2018counting, shen2024identification, kulits2020automated}, and trajectory tracking~\cite{kang2011semiautomated,folkesson2007feature}. With advancements in sonar technology and AI-empowered data processing, the demand for real-time sonar data analysis has significantly increased, particularly in time-sensitive applications such as environmental anomaly detection and in-season fishery management. In these contexts, timely and accurate sonar analytics can greatly enhance fast response efforts, enabling more effective decision-making.

\begin{figure}
  \centering
  \begin{subfigure}[b]{0.36\linewidth}
    \includegraphics[width=\linewidth]{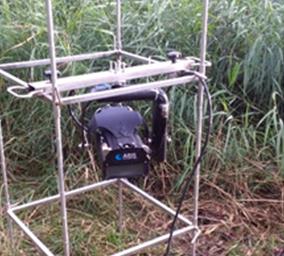}
    \caption{}
    \label{fig:aris-holder}
  \end{subfigure}
  \hspace{0.02\linewidth}
  \begin{subfigure}[b]{0.50\linewidth}
    \includegraphics[width=\linewidth]{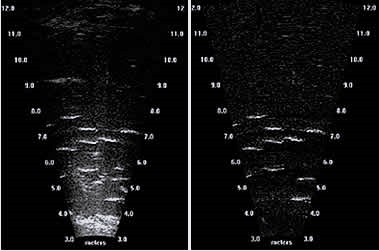}
    \caption{}
    \label{fig:aris-frame}
  \end{subfigure}
    \vspace{-0.2cm}
  \captionsetup{width=0.9\linewidth}
  \caption{\textbf{(a)} a mounted ARIS sonar radar, \textbf{(b)} sample frames from ARIS sonar.}
  \label{fig:sample}
  \vspace{-0.3cm}
\end{figure}

Despite decades of research, achieving live and accurate sonar analytics remains challenging due to several factors. Sonar radars are often deployed in environments where optical cameras are ineffective due to low or zero visibility, such as underwater. As shown in Figure~\ref{fig:sample}, objects such as fish, wild animals, and intruders frequently appear small and blurry in sonar frames, complicating typical analytics tasks~\cite{abu2019enhanced}. Additionally, sonar data typically exhibit limited diversity and suffer from significant noise, with insufficient texture details necessary for effectively training and fine-tuning deep neural networks (DNNs)~\cite{li2019underwater}.

Another challenge is balancing data fidelity and freshness in live sonar analytics~\cite{edgednn, dnnoffloading, wang2020joint}. Sonar data is rich in information by nature. Unlike modern embedded cameras with built-in processing capabilities~\cite{zeng2020distream}, off-the-shelf sonar radars generate raw echo frames and basic sensory data, requiring reliable edge infrastructure for further processing. Without efficient handling, this results in substantial overhead and degrades overall analytics performance.

\begin{figure}
  \includegraphics[width=0.4\textwidth]{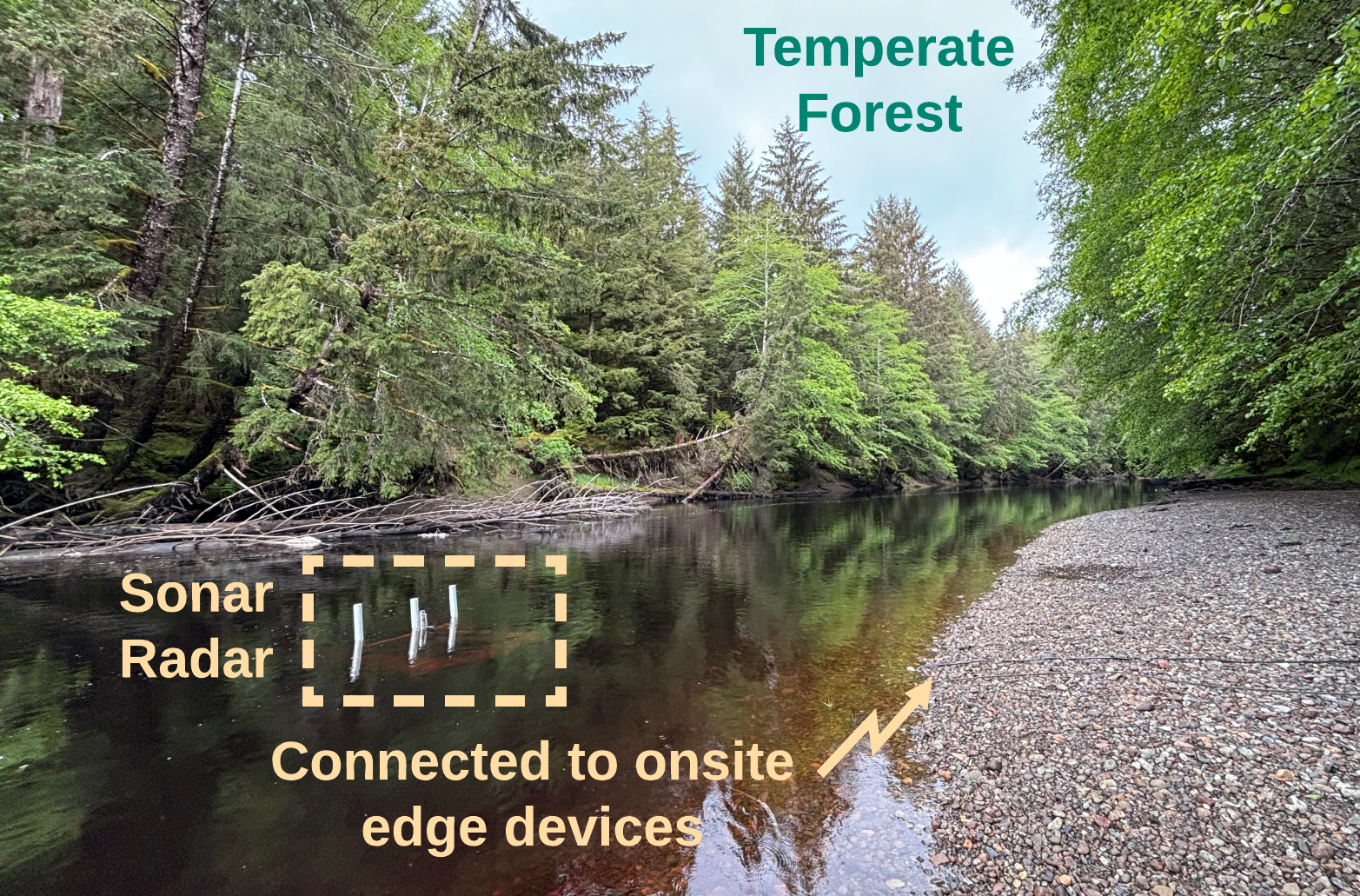}
  \captionsetup{width=0.9\linewidth}
  \caption{Sonar radar and edge devices deployed in off-grid temperate forest.}
  \label{fig:deploy}
  \vspace{-0.3cm}
\end{figure}

In addition, unlike urban surveillance scenarios with robust network infrastructure, sonar radar is often deployed in wild ecosystems with limited network coverage and energy resources. Field studies in North America showed that individuals and agencies rely on satellite internet providers, such as SpaceX's Starlink~\cite{fang2024streaming, zhao2023realtime}, for accessing computing power to process streamed sonar data. However, Starlink's peak power consumption of up to 200 Watts is double that of the sonar radar system's 100 Watts. In these areas, sustainable energy sources such as solar power are typically the only viable option. As shown in Figure~\ref{fig:deploy}, the temperate forest environment exacerbates the challenge, as changing weather conditions, sudden storms, and cloud cover significantly impact energy availability. Therefore, unique energy and network characteristics must be considered when designing sustainable and enduring solutions for live sonar analytics.

In this work, we pose the following question: \textit{"Can a live sonar streaming and analytics system be developed to operate effectively under the extreme constraints of wild ecosystems?"} We anticipate that an affirmative answer will offer insights for the long-term operation of similar systems on processing generalized acoustic data. 

To address this question, we developed SALINA for \textbf{S}ust\textbf{A}inable \textbf{LI}ve so\textbf{N}ar \textbf{A}nalytics, in close collaboration with a multidisciplinary team of biologists, forest technologists, and electricians. SALINA's novelty lies in three key aspects: \textit{sonar data channel population, DNN model enhancement with spatial-temporal feature adaptations}, and \textit{sustainable sonar streaming and energy planning}. The sonar data channel population ensures robust data perception of both human and machine intelligence, while the DNN model enhancement incorporates spatial-temporal feature adaptations to further improve analytics performance in challenging conditions. Additionally, edge-cloud collaboration is leveraged during inference to balance accuracy and power efficiency. Furthermore, the energy planning module is designed to handle volatile weather patterns, such as sudden storms and fluctuating cloud cover, ensuring continuous operation while maintaining resource efficiency in rapidly changing environments.



SALINA was deployed for six months at two inland river sites in British Columbia, Canada, located deep within temperate forests and operating entirely off-grid, relying solely on sustainable solar power. It provided continuous 24/7 underwater monitoring to support the strategic planning of the First Nations’ fishery stewardship and wildlife restoration efforts. The analytics results also contributed to biological research on tracking North American Atlantic salmon. Extensive real-world experiments showed SALINA's superiority, with up to 9.5\% improvement in average precision and 10.1\% improvements in tracking metrics when monitoring underwater objects. The energy planning module effectively schedules energy usage, preventing system outages due to extreme weather conditions and saving considerable contingency costs. Our contributions can be summarized as follows.

\begin{itemize}
\item{We developed the first known system for real-time processing and analysis of acoustic sonar data, incorporating edge and cloud collaboration for live analytics in wild ecosystems.}
\item{We addressed unique challenges in sonar analytics by constructing a novel channel population pipeline to resolve issues such as acoustic shadow and reverberation, improving detection and tracking accuracy.}
\item{Through reconstructing data channels and finetuning pre-trained DNN models, we observed improved detection and tracking performance, compared to state-of-the-art methods~\cite{kay2022caltech} on three benchmarking datasets.}
\item{SALINA's sustainable sonar streaming and energy planning module is explicitly designed to withstand volatile weather patterns, such as sudden storms and fluctuating cloud cover, ensuring continuous operation and resource efficiency in rapidly changing environments.}
\item{The sonar dataset used in this study has been organized and released for community use, with the potential to generate new insights and discoveries that benefit society.}
\end{itemize}


The remainder of this paper is outlined as follows:

Section~\ref{sec:background} presents comprehensive backgrounds and research motivations, discussing the sonar analytics characteristics and challenges; Section~\ref{sec:overview} introduces the SALINA architecture and its design overview; Section~\ref{sec:edgepre} details the data preprocessing and wrangling techniques; Section~\ref{sec:dnn} describes the DNN model adaptations for both on-premise and cloud inference; Section~\ref{sec:energy} explains the sonar streaming and energy planning design; Section~\ref{sec:eval} provides system evaluations; Section~\ref{sec:discussion} offers further discussion and Section~\ref{sec:conclusion} concludes the paper.
\begin{figure*}
  \includegraphics[width=0.99\textwidth]{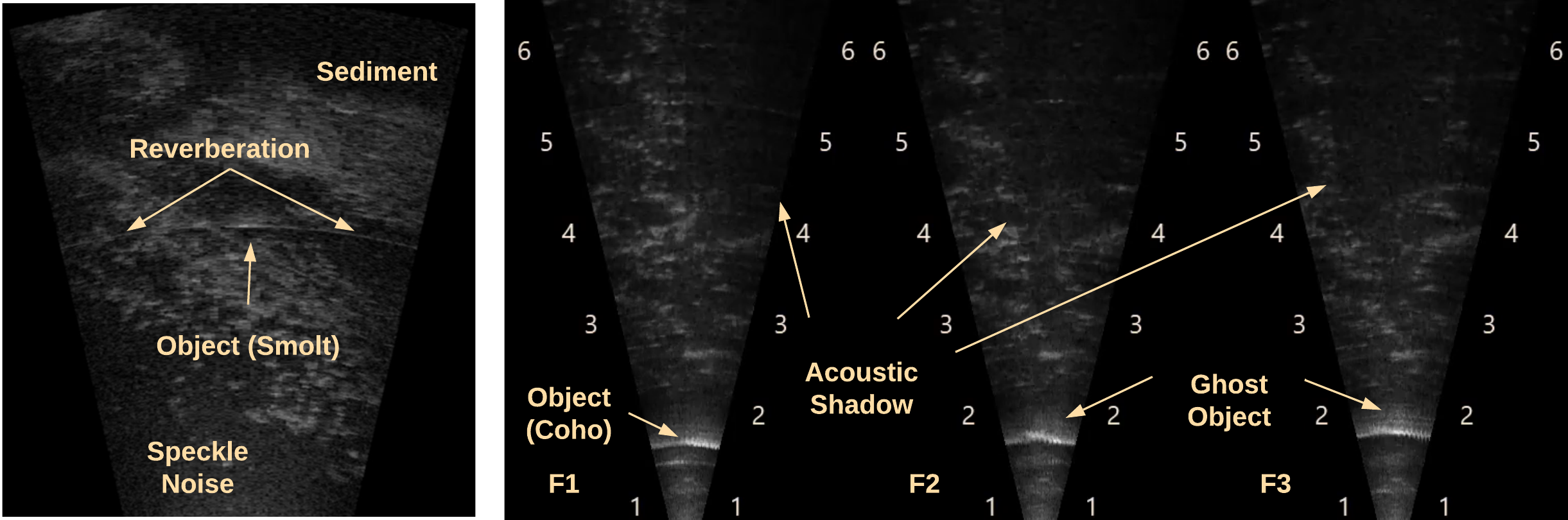}
    \vspace{-0.1cm}
  \caption{Unique observations in sonar frames. The left image shows reverberation effects (elongated curves) and pervasive speckle noise. The right images display three consecutive frames (F1, F2, and F3) illustrating the moving acoustic shadow caused by a coho salmon. The ghost objects are shown as stacked highlighted areas above the fish.}
  \label{fig:unique}
  \vspace{-0.3cm}
\end{figure*}

\section{Background and Motivations}
\label{sec:background}
In this section, we introduce the background, opportunities, and challenges in developing live sonar analytics in wild ecosystems.

\subsection{Sonar Frame Capture}

In this work, we focus primarily on data analytics of multi-beam sonar. In contrast to passive sonar and side scan sonar~\cite{williams2015fast}, multi-beam sonar works by emitting multiple beams of sound waves from a transducer, which is mounted on a fixed rack or towed behind a boat. Each sonar frame visualizes the reflected intensity of these beams, providing clear underwater imagery even in low-visibility conditions. This enables effective detection of aquatic life, mapping of structures, and monitoring of submerged assets, making it suitable for real-time underwater exploration and analysis. As shown in Figures ~\ref{fig:aris-frame} and ~\ref{fig:unique}, the frames are typically presented in grayscale or false-color, where lighter areas indicate stronger reflections or denser objects, and darker areas represent weaker reflections or softer, less dense materials.

\subsection{Characteristics of Live Sonar Analytics}
\textbf{Emerging demand for real-time feedback.} Advancements in sonar technology and AI-driven data processing have created a growing need for real-time feedback in applications such as environmental anomaly detection and in-season fishery management, where timely insights are crucial for effective decision-making. For example, during the salmon return season, real-time sonar monitoring enables fishery programs to track populations, assess environmental impacts, and adjust operations promptly, ensuring compliance with regulations and sustainable harvests. Without live sonar data, decisions would rely on delayed or less accurate information, potentially leading to poor management and environmental risks. Despite these benefits, achieving both high data fidelity and freshness in live sonar analytics remains challenging~\cite{edgednn, dnnoffloading, wang2020joint}. Beyond fishery management, live sonar analytics can support intrusion detection and disaster management, providing precise and timely insights to adapt to diverse scenarios and dynamics.

\textbf{Lack of quality dataset and pre-trained models.} Compared to conventional computer vision detection and tracking (D\&T) tasks, identifying and continuously tracing objects in sonar data is inherently challenging. Objects in real-world scenarios, such as fish, wild animals, and intruders, often appear small and blurry in echo frames~\cite{abu2019enhanced}. In addition, underwater datasets typically have limited diversity, and real-world samples are accompanied by heterogeneous noise and lack texture details~\cite{li2019underwater}, making it difficult to train and fine-tune DNN models for D\&T tasks. Furthermore, there are only a few publicly available datasets for pre-training, partly due to the limited number of research teams focusing on this area.

\textbf{Noises cause detection loss.} Compared with environmental noise, gaussian noise, and motion blur which are commonly seen and handled in conventional video analytics systems, speckle noise is a type of granular noise that frequently appears in images or signals acquired through coherent imaging systems such as sonar radar. It arises due to interference between coherent wavefronts and varies with the relative positions of the object and sensor. As shown in Figure~\ref{fig:unique}, it manifests as random fluctuations in brightness or intensity, creating a grainy or speckled appearance. In sonar analytics, such noises can adversely affect D\&T tasks, leading to a loss of accuracy or degraded performance.

\textbf{Acoustic shadows and reverberations.}
Additionally, acoustic shadows and reverberations caused by sound waves can further complicate object observation by obscuring object presence. For example, when placed underwater, the interaction of sound waves bouncing back and forth between the river bottom and the water surface before reaching the transducer generates multiple shifting bottom images ~\cite{kay2022caltech}. Consequently, this type of echo may produce multiple images of individual objects, appearing offset from each other, often referred to as "ghost objects". A typical example is presented in Figure~\ref{fig:unique}, showing a moving acoustic shadow and ghost objects caused by a Coho salmon in sonar frames. 

\begin{figure}
  \includegraphics[width=0.40\textwidth]{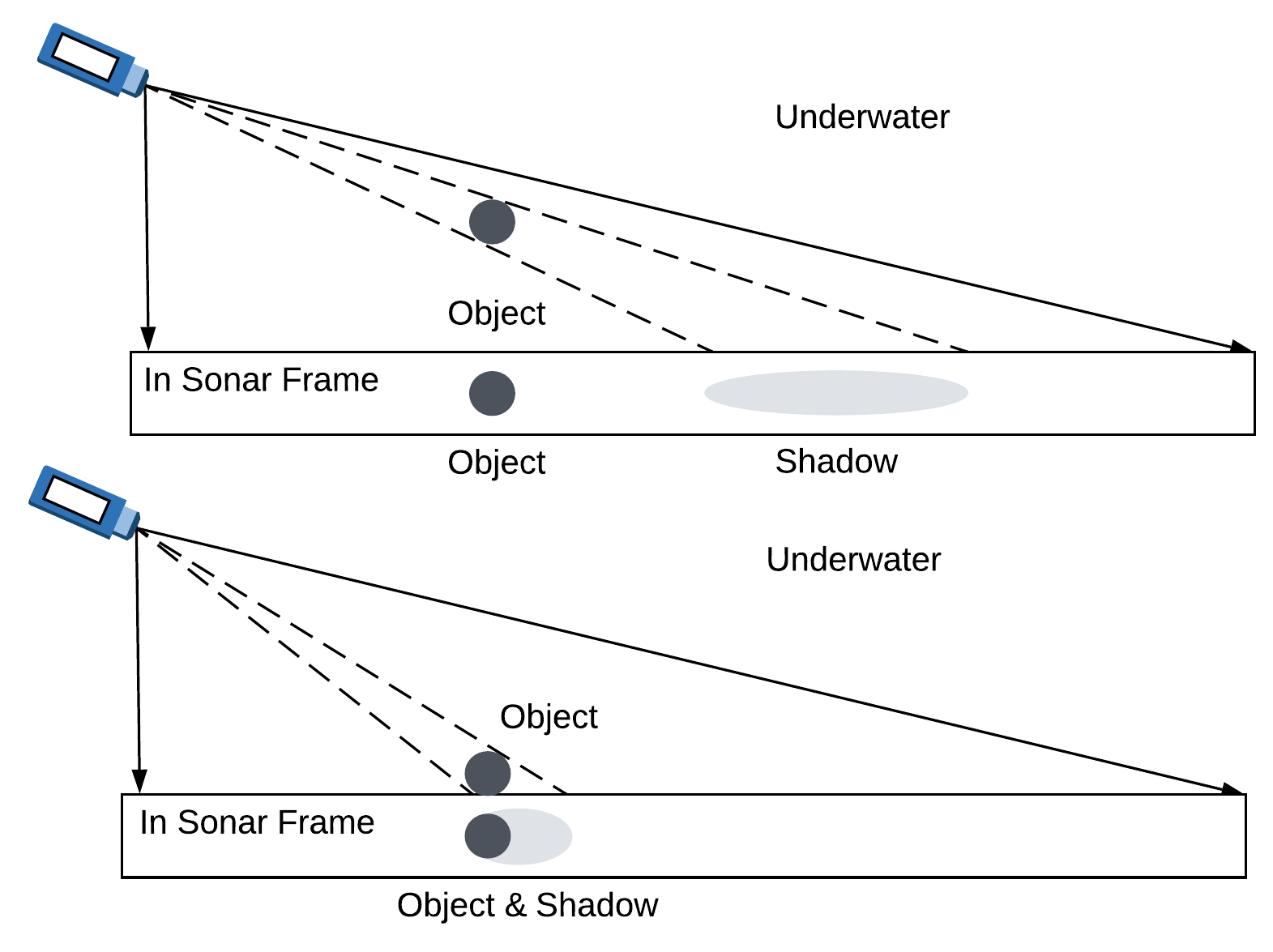}
    \vspace{-0.1cm}
  \caption{Acoustic shadows in sonar frames.}
  \label{fig:shadow}
  \vspace{-0.3cm}
\end{figure}

Furthermore, we also illustrate the form of acoustic shadows in Figure~\ref{fig:shadow}. In this figure, the objects closer to the sonar radar have larger acoustic shadows in the sonar frame, while those farther from the radar and closer to the bottom have smaller shadows. Such acoustic shadows can potentially interfere with the detection and tracking of other objects. Note that the presence and intensity of these ghost objects can vary based on factors such as the river's shape, sonar radar orientation, and water level. These factors determine how much these visual anomalies distort the perceived location and size of objects in the sonar frame, potentially leading to inaccurate object detection or misclassification. To improve monitoring and research accuracy, it is crucial to quantify the impact of these acoustic properties and develop methods to mitigate their effects on visual data interpretation.

\subsection{Deployment Challenges}
\textbf{Limited compute power and energy constraints at the first mile.} Sonar data contains a vast amount of information that needs to be processed in real time. In contrast to modern embedded IP cameras that equip powerful processing units, off-the-shelf sonar radar only outputs raw echo frames and other basic sensory status data. This necessitates a reliable edge infrastructure to handle data pre-processing and wrangling. Possible edge designs include deploying lightweight models on low-power devices, offloading computationally intensive tasks to cloud servers, and employing data compression techniques to reduce transmission costs. Without these optimizations, processing overheads can degrade overall system performance. Note in wild environments, sustainable energy sources such as solar power are typically the only viable option. The changing weather conditions, sudden storms, and cloud cover significantly impact energy availability, posing challenges in continuous operation and maintenance. 

\textbf{Thin network coverage in the wild.} Unlike urban surveillance scenarios~\cite{zhang2018awstream,jia2023rdladder}, which benefit from well-established network infrastructure, sonar radar is often deployed in wild environments with limited network coverage~\cite{ma2023network}. Geographic factors such as mountains, dense forests, or other natural obstacles can obstruct network signal transmission, further reducing coverage. In these areas, individuals and agencies often rely on satellite internet providers, such as SpaceX's Starlink, for connectivity. Therefore, these unique network constraints must be carefully considered when designing and developing a live sonar analytics system.

\begin{figure}
  \includegraphics[width=0.45\textwidth]{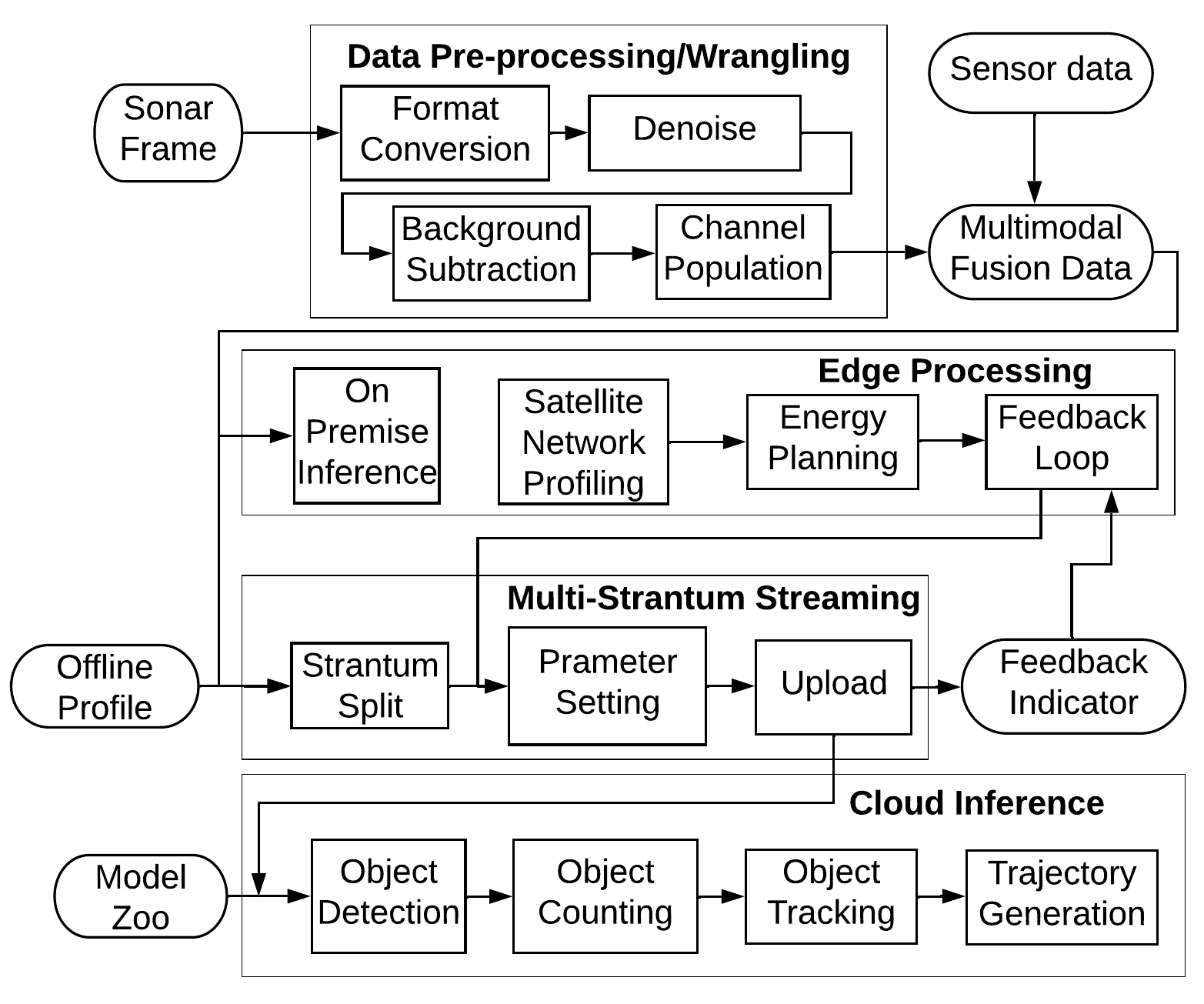}
    \vspace{-0.1cm}
  \caption{Module design.}
  \label{fig:module}
  \vspace{-0.4cm}
\end{figure}

\section{SALINA Overview}
\label{sec:overview}
Considering the unique characteristics and challenges in sonar analytics, we present SALINA, a sustainable live sonar analytics system. The major modules of SALINA are presented in Figure~\ref{fig:module}, which we will detail further. 

\subsection{Data Preprocessing and Wrangling}

In this work, we employ advanced data preprocessing and wrangling techniques to enhance live sonar analytics. By converting acoustic intensity measurements to a grayscale frame, we facilitate the use of various image processing methods to populate data channels for deep neural network (DNN) models to work on downstream tasks. For example, for real-time background subtraction, we utilize the Mixture of Gaussians (MOG) method~\cite{stauffer1999adaptive, zivkovic2004improved}, which models the background using multiple Gaussian distributions per pixel. This approach adapts dynamically to scene changes, ensuring efficiency and robustness. 

Additionally, we propose a novel hedging design to utilize background removal data by creating three-channel outputs compatible with conventional RGB formats. The first channel retains the original sonar frame, while the second and third channels apply guided filtering ~\cite{he2010guided} with different guidance images. This strategy improves the differentiation of noises and acoustic shadows, enhancing the performance of detection and tracking tasks. By integrating these preprocessing steps, our methodology ensures real-time processing capabilities and significantly improves the accuracy of downstream analytics. Further details are presented in Section 4.

\subsection{Sonar Dataset Preparation}
Due to the scarcity of public underwater sonar datasets, we developed custom datasets for training, fine-tuning, and benchmarking purposes. In sonar analytics scenarios, visual similarities between objects make it challenging to distinguish them based on appearance alone, necessitating the use of motion and behavioral patterns for effective tracking. This complexity is compounded by factors such as intricate backgrounds, inconsistent frame-to-frame visibility, and varying numbers of objects in the scene—issues that are less common in controlled laboratory tracking studies. To address these challenges, we hired a third-party annotation service to collect multiple object-tracking annotations for all objects in the converted sonar clips in grayscale. The objects include salmon, otter, and smolt. Annotators were provided with the sonar frames and instructed to tightly box all visible objects of interest using v7 software~\cite{v7labs}.

The dataset was curated from three months of sonar clips obtained from two distinct rivers. For the YK River, the dataset contains 454,941 bounding boxes across 196,937 frames from 1,346 clips. This dataset was split into training and validation sets with an 80/20 ratio. For the more challenging KN River, the training set comprises 132,010 bounding boxes across 162,681 frames from 483 clips. The validation set consists of 18,551 bounding boxes across 30,521 frames from 65 clips.

\subsection{DNN Model Adaptations}
Section 5 explores DNN model adaptations for both on-premise and cloud inference to generate sonar analytics results. The need for both on-premise and cloud inference arises from the distinct advantages each offers. On-premise inference offers edge-only processing with immediate feedback, a low energy profile, and resilience to occasional network outages. In contrast, cloud inference provides greater computational power, enabling complex and resource-intensive analyses with higher accuracy, assuming stable energy and network are available for data transmission to the cloud. Leveraging both on-premise and cloud inference also addresses the heterogeneous settings of different sonar sites, where device capabilities, energy availability, and network conditions vary significantly in the wild. 

We adopt Convolutional Neural Networks (CNN) for on-premise inference, leveraging their modularity and anchor-free architecture to reduce false positives in sonar data. 
For cloud inference, we adapt the deformable detection transformer (DETR) ~\cite{zhu2021deformable} as our still sonar frame detector. We simplify its design by excluding multi-scale feature representations and concentrating on the last stage of the backbone. The modified detector for sonar frames incorporates a spatial transformer encoder and decoder, which encode each frame into spatial object queries and memory encodings. Additionally, we introduce a temporal transformer to enhance object detection by leveraging temporal information between adjacent sonar frames, performing co-attention between online queries and temporally aggregated features. We refer to these adaptations for sonar data as the Spatial-Temporal Sonar Vision Transformer (STSVT).

\subsection{Sustainable Sonar Streaming}
In Section 6, we introduce a joint design for sonar streaming and energy planning. This integration is necessary because supporting cloud inference for better analytics accuracy requires sonar streaming over satellite networks, which consumes the majority of power in the entire system. Our design considers both weather-affected Starlink connectivity and solar photovoltaic (PV) energy production. We observe that Starlink's throughput is significantly affected by precipitation, with an average 15\% drop during precipitation and notable reductions during heavy rain due to rain attenuation affecting Ka- and Ku-band radio waves. Similarly, solar PV energy production is highly sensitive to weather conditions. We leverage a modified U-Net architecture for short-term solar PV output forecast, incorporating residual blocks and pruning to enhance efficiency. Our multi-stratum streaming optimization integrates satellite communication performance metrics and PV energy forecasts to dynamically adjust streaming configurations and data rates, ensuring Pareto-optimality even in volatile weather conditions.

\section{Sonar Data Preprocessing and Wrangling}
\label{sec:edgepre}

\subsection{Denoising and Background Removal}
As a default approach in sonar analytics, converting acoustic intensity measurements to a grayscale frame facilitates the use of image processing and computer vision techniques. This conversion results in a single-channel image representation, where each pixel value corresponds to the intensity of the sonar return at that location. Such a representation is essential for leveraging existing DNN-based detection and tracking models, which are typically designed for processing visual data.

A common approach for preprocessing these grayscale sonar images involves the application of Gaussian blur~\cite{forsyth2002computer}. Gaussian blur smooths the image by reducing high-frequency noise, which is often prevalent in sonar data due to various environmental and sensor-related factors. This process helps in enhancing the detection of significant features and objects by suppressing noise that might otherwise lead to performance degradation in downstream tasks.

In the context of background subtraction, state-of-the-art approaches~\cite{kay2022caltech} involve computing the average of all frames in the sonar clips and subtracting this average from each frame. This approach, known as frame differencing with mean subtraction, effectively highlights moving objects while suppressing static background elements. However, this method has notable limitations when applied to real-time settings. The requirement to maintain and process the entire sonar clips to compute the average frame is computationally intensive and may not be feasible in scenarios where real-time processing is essential. 

In our work, we propose several real-time steps to address these challenges. First, we use the Mixture of Gaussians (MOG) method~\cite{stauffer1999adaptive} to differentiate between foreground and background in sonar frames. MOG is computationally efficient, dynamically updates with each frame, and can handle speckle noise, reverberations, and acoustic shadows by modeling each pixel as a mixture of Gaussian distributions. This allows it to separate random fluctuations from the stable background and adapt to the varying patterns of reverberations and shadows, minimizing their impact on detection performance.
Mathematically, the MOG model is formulated as follows:

\[
p(x) = \sum_{k=1}^{K} \pi_k \mathcal{N}(x \mid \mu_k, \Sigma_k)
\]

where \( p(x) \) is the probability of a pixel value \( x \), \( K \) is the number of Gaussian components, \( \pi_k \) are the mixing coefficients with \( \sum_{k=1}^{K} \pi_k = 1 \) and \( \pi_k \geq 0 \), and \( \mathcal{N}(x \mid \mu_k, \Sigma_k) \) represents the Gaussian distribution with mean \( \mu_k \) and covariance \( \Sigma_k \). The parameters \( \mu_k \), \( \Sigma_k \), and \( \pi_k \) are continuously updated based on incoming pixel values, allowing the model to adapt to new background conditions dynamically. 
The update steps for the parameters are as follows:

1. Mean (\( \mu_k \)):
   \[
   \mu_{k}^{(t+1)} = (1 - \alpha) \mu_{k}^{(t)} + \alpha x_t
   \]

2. Covariance (\( \Sigma_k \)):
   \[
   \Sigma_{k}^{(t+1)} = (1 - \alpha) \Sigma_{k}^{(t)} + \alpha (x_t - \mu_{k}^{(t+1)})(x_t - \mu_{k}^{(t+1)})^T
   \]

3. Weight (\( \pi_k \)):
   \[
   \pi_{k}^{(t+1)} = (1 - \alpha) \pi_{k}^{(t)} + \alpha M_k
   \]

Herein, \( \alpha \) is the learning rate, \( x_t \) is the current pixel value at time \( t \), and \( M_k \) is an indicator function that equals 1 for the matched Gaussian \( k \) and 0 otherwise.

\begin{figure}
  \centering
  \begin{subfigure}[b]{0.35\linewidth}
    \includegraphics[width=\linewidth]{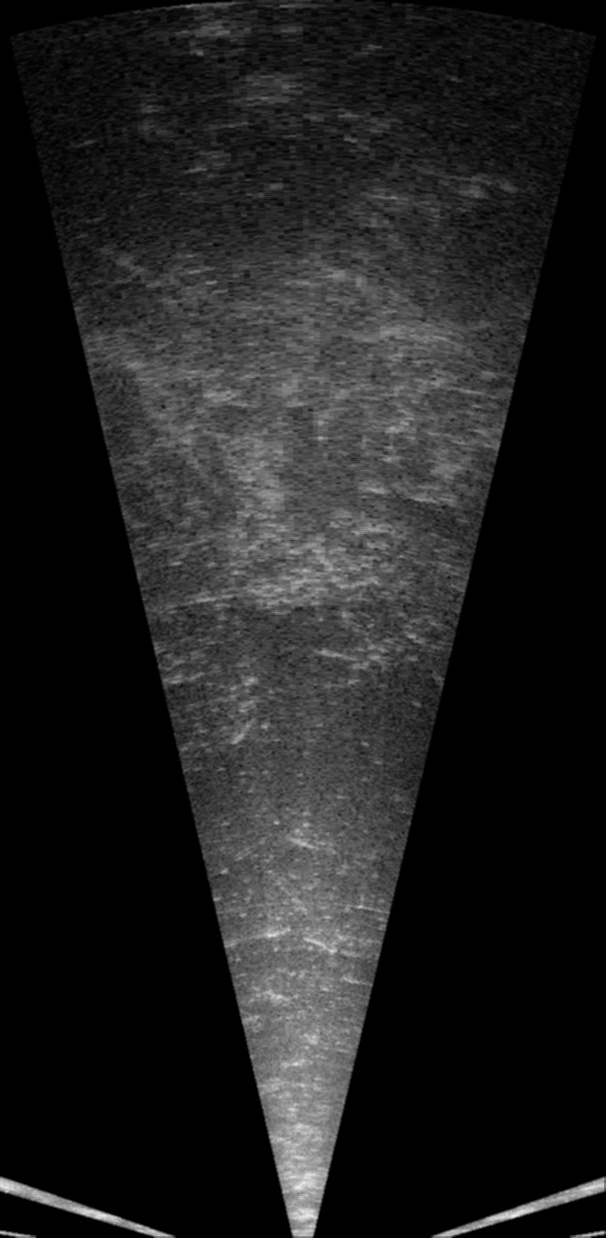}
    \caption{}
    \label{fig:ch-ori}
  \end{subfigure}
  \hspace{0.05\linewidth}
  \begin{subfigure}[b]{0.35\linewidth}
    \includegraphics[width=\linewidth]{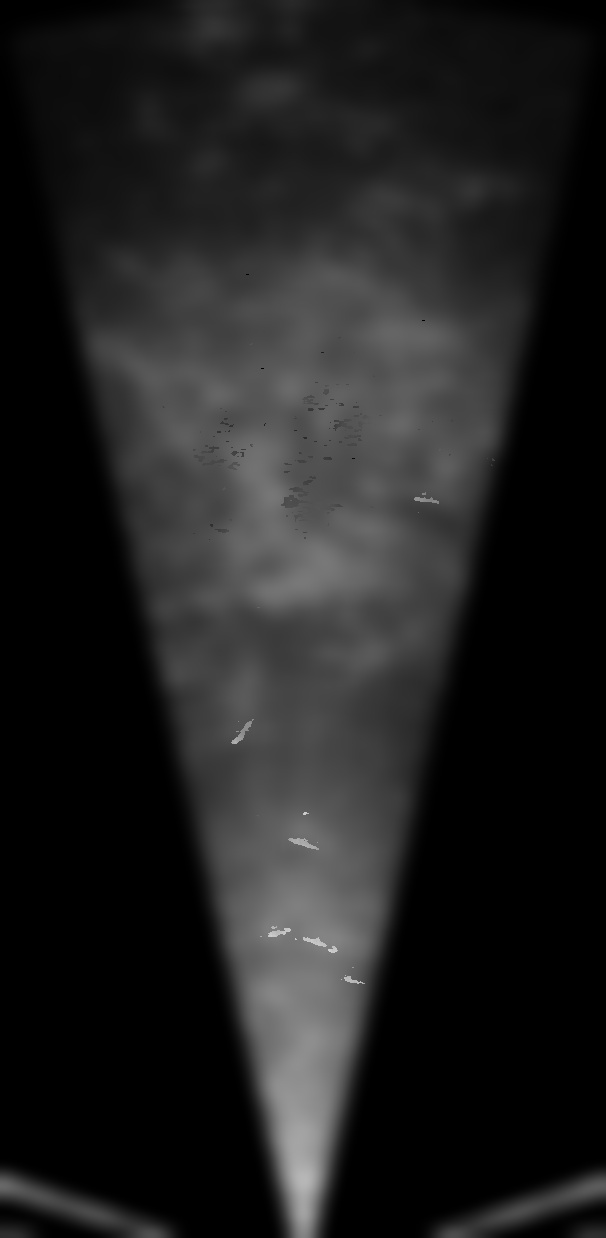}
    \caption{}
    \label{fig:ch-guid-img}
  \end{subfigure}
  \\
  \begin{subfigure}[b]{0.35\linewidth}
    \includegraphics[width=\linewidth]{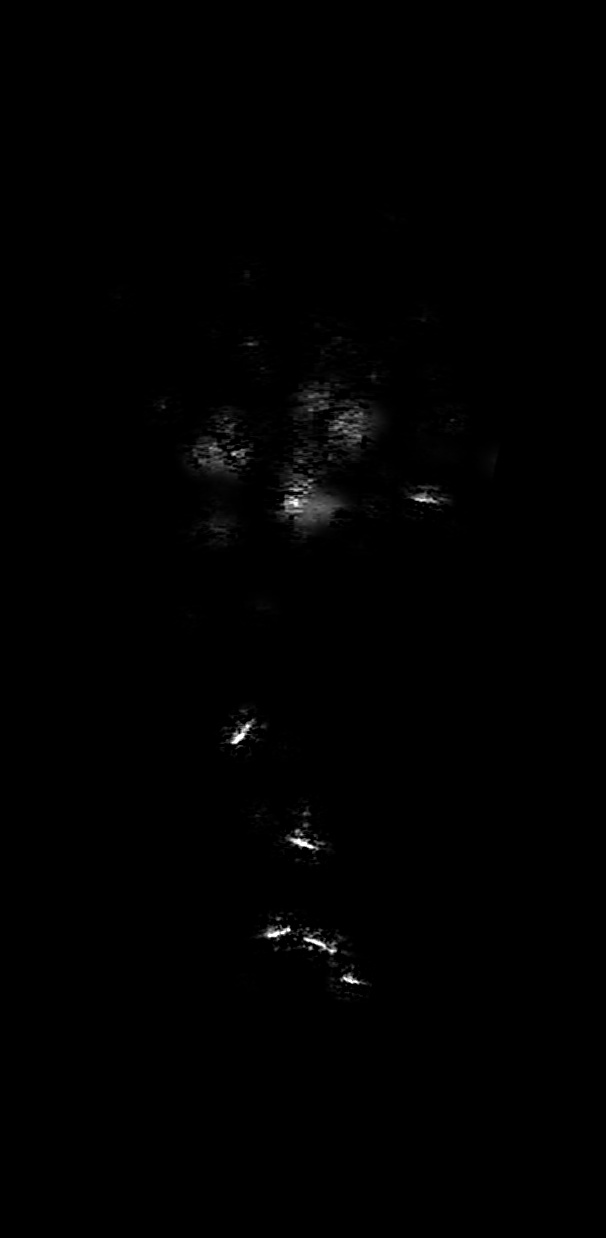}
    \caption{}
    \label{fig:ch-guid-mog}
  \end{subfigure}
  \hspace{0.05\linewidth}
  \begin{subfigure}[b]{0.35\linewidth}
    \includegraphics[width=\linewidth]{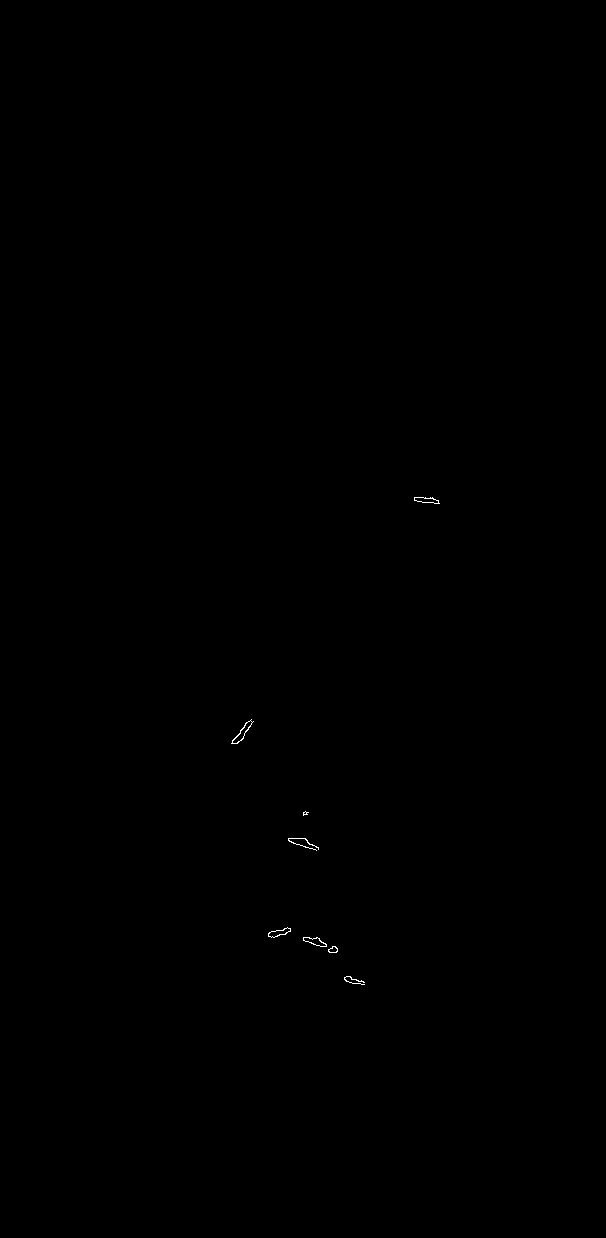}
    \caption{}
    \label{fig:ch-guid-img-edge}
  \end{subfigure}
  \vspace{-0.1cm}
  \captionsetup{width=0.95\linewidth}
  \caption{Channels: (a) original, (b) guided filtered, (c) guided MOG, (d) edge preservation after applying Canny edge detector~\cite{canny1986computational}.}
  \label{fig:channels}
  \vspace{-0.3cm}
  \end{figure}

\begin{figure}
  \begin{subfigure}[b]{0.35\linewidth}
    \includegraphics[width=\linewidth]{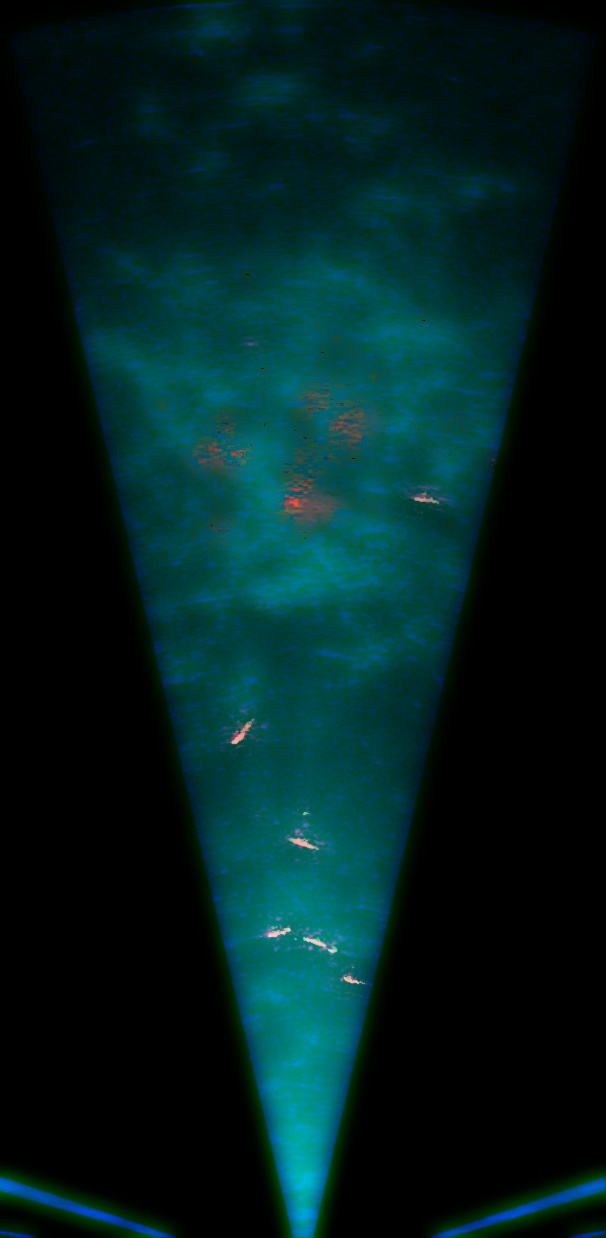}
    \caption{}
    \label{fig:ch-pop}
  \end{subfigure}
    \hspace{0.05\linewidth}
  \begin{subfigure}[b]{0.35\linewidth}
    \includegraphics[width=\linewidth]{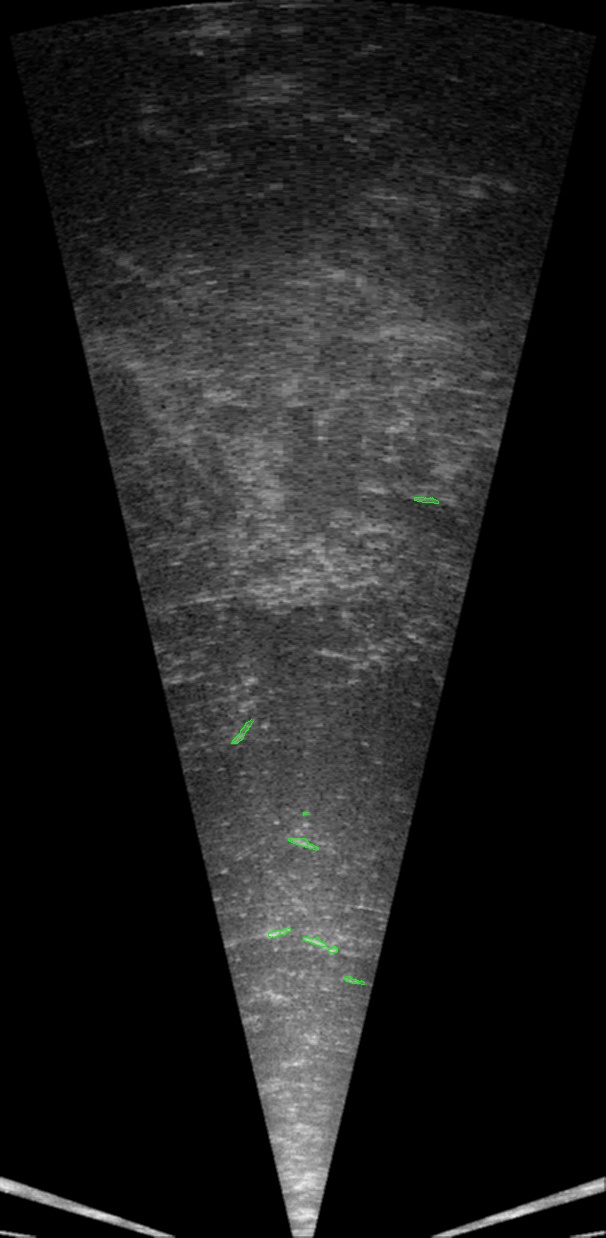}
    \caption{}
    \label{fig:ch-mot}
  \end{subfigure}
  \vspace{-0.1cm}
  \captionsetup{width=0.95\linewidth}
  \caption{Channel population: (a) visualizing converted three channels, (b) motion detection results.}
  \label{fig:population}
  \vspace{-0.3cm}
\end{figure}

In addition, the MOG approach is particularly well-suited for our context, as it effectively handles the dynamic nature of sonar data, where background conditions change rapidly due to factors such as water currents or the movement of mobile organisms. By continuously updating the Gaussian parameters, MOG adapts to these variations, ensuring an accurate background model and enabling robust foreground differentiation. 


\subsection{Hedging Design in Channel Population}
Furthermore, we propose a novel hedging design to effectively utilize background removal data from MOG outputs for sonar analytics. Leveraging original sonar frames and MOG outputs, this design populates three data channels to fit the RGB format used by existing DNN detection models. We show an illustration example in Figure~\ref{fig:channels}. The first channel (Figure~\ref{fig:ch-ori}) is the original sonar frame, preserving the raw intensity data. The second and third channels (Figures~\ref{fig:ch-guid-img} and \ref{fig:ch-guid-mog}) undergo guided filtering, but with different guiding images, to enhance the differentiation between noises and acoustic shadows.

The guided filter~\cite{he2010guided} is an edge-preserving smoothing technique that leverages a guidance image to perform filtering. Mathematically, the guided filter can be defined as follows:
Given an input image \( I \) and a guidance image \( G \), the output image \( Q \) is computed by minimizing the following cost function for each window \( \omega_j \) centered at pixel \( j \):

\[
E(a_j, b_j) = \sum_{i \in \omega_j} \left( (I_i - (a_j G_i + b_j))^2 + \epsilon a_j^2 \right)
\]

where \( a_j \) and \( b_j \) are the linear coefficients that fit the guidance image \( G \) to the input image \( I \) within the window \( \omega_j \), and \( \epsilon \) is a regularization parameter to prevent overfitting.

The solution to this minimization problem yields the coefficients:

\[
a_j = \frac{\frac{1}{|\omega_j|} \sum_{i \in \omega_j} G_i I_i - \mu_k \bar{I}_j}{\sigma_j^2 + \epsilon}
\]

\[
b_j = \bar{I}_j - a_j \mu_j
\]

where \( \mu_j \) and \( \sigma_j^2 \) are the mean and variance of \( G \) in the window \( \omega_j \), and \( \bar{I}_j \) is the mean of \( I \) in the same window. The output image \( Q \) is then computed as:

\[
Q_i = a_j G_i + b_j
\]

In our design, the second channel uses the original sonar frame as the input image 
$I$ and the MOG foreground result as the guidance image $G$. This configuration helps to highlight features that are consistent with the foreground, reducing noise and emphasizing significant sonar returns.

The third channel reverses this relationship: the MOG foreground result becomes the input image $I$, and the original sonar frame serves as the guidance image $G$. This setup enhances features that differ from the foreground model, effectively isolating moving objects and suppressing static elements.

This hedging strategy leverages the strengths of both configurations. The first guided filter (original as input, MOG as guide) emphasizes foreground consistency, which helps to suppress noise and highlight significant features. The second guided filter (MOG foreground as input, original as guide) isolates moving objects, differentiating them from static background elements. Together, these channels preserve the edges of objects (Figure~\ref{fig:ch-guid-img-edge}) and provide a robust representation (Figure~\ref{fig:ch-pop}) that is well-suited for DNN-based detection models. We further propose a motion detection solution by using the converted three channels as input for Canny edge detection. This approach, performed in real time, serves as an effective motion detector that is resilient to noise. The motion detector helps to reduce the number of frames that need to be processed in subsequent stages, significantly reducing the processing and transmission load. An example of the identified prominent edges can be seen in Figure~\ref{fig:ch-mot}.

\section{DNN Model Adaptation}
\label{sec:dnn}

\subsection{On Premise Inference with CNN}

Using Convolutional Neural Networks (CNN) for sonar data detection provides a robust solution for real-time analysis. As suggested in~\cite{kay2022caltech}, such models as YOLO~\cite{yolov8}, Faster R-CNN~\cite{ren2015faster}, and SSD~\cite{liu2016ssd}, designed for rapid object detection, align well with the demands of sonar applications, where swift and accurate identification of objects is critical. These models' inherent capability to handle various data types ensures efficient processing of sonar signals and reliable detection results once adapted.
\begin{figure}
  \includegraphics[width=0.49\textwidth]{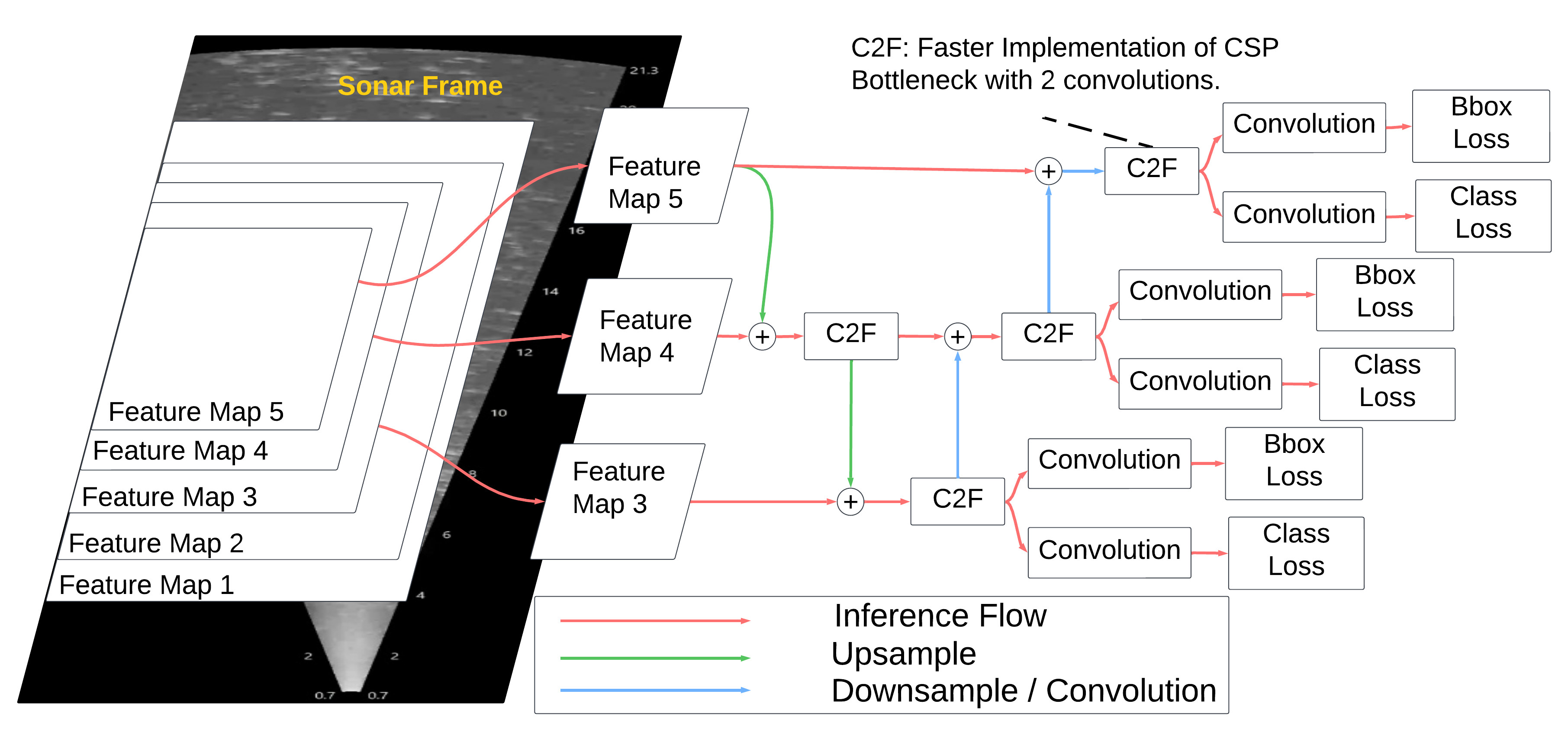}
    \captionsetup{belowskip=-10pt} 

  \caption{DNN architecture for on premise inference.}
    \vspace{-0.1cm}
  \label{fig:cnn}
\end{figure}

We adopt YOLOv8 as our base model for on-premise inference, using our converted 3-channel input. As shown in Figure 8, YOLOv8 integrates features from previous versions, including Multi-Scale Predictions (v3), PANet (v4), and the Efficient Layer Aggregation Network (ELAN) (v7), while introducing the c2f block for improved feature extraction and aggregation. Its anchor-free architecture reduces candidate bounding boxes, effectively minimizing false positives in object detection for live sonar analytics.

 
Additionally, the lightweight architecture of YOLOv8 ensures low latency and rapid inference, making it ideal for deployment on edge devices such as the Jetson Orin Nano~\cite{nvidia_jetson_orin_nano}. These devices are optimized for edge AI applications, offering the necessary computational power while maintaining low energy consumption and thermal efficiency. This makes YOLOv8 an excellent choice for on-premise inference, offering immediate feedback, low energy usage, and resilience to occasional network outages in the wild.

\begin{figure}
  \includegraphics[width=0.35\textwidth]{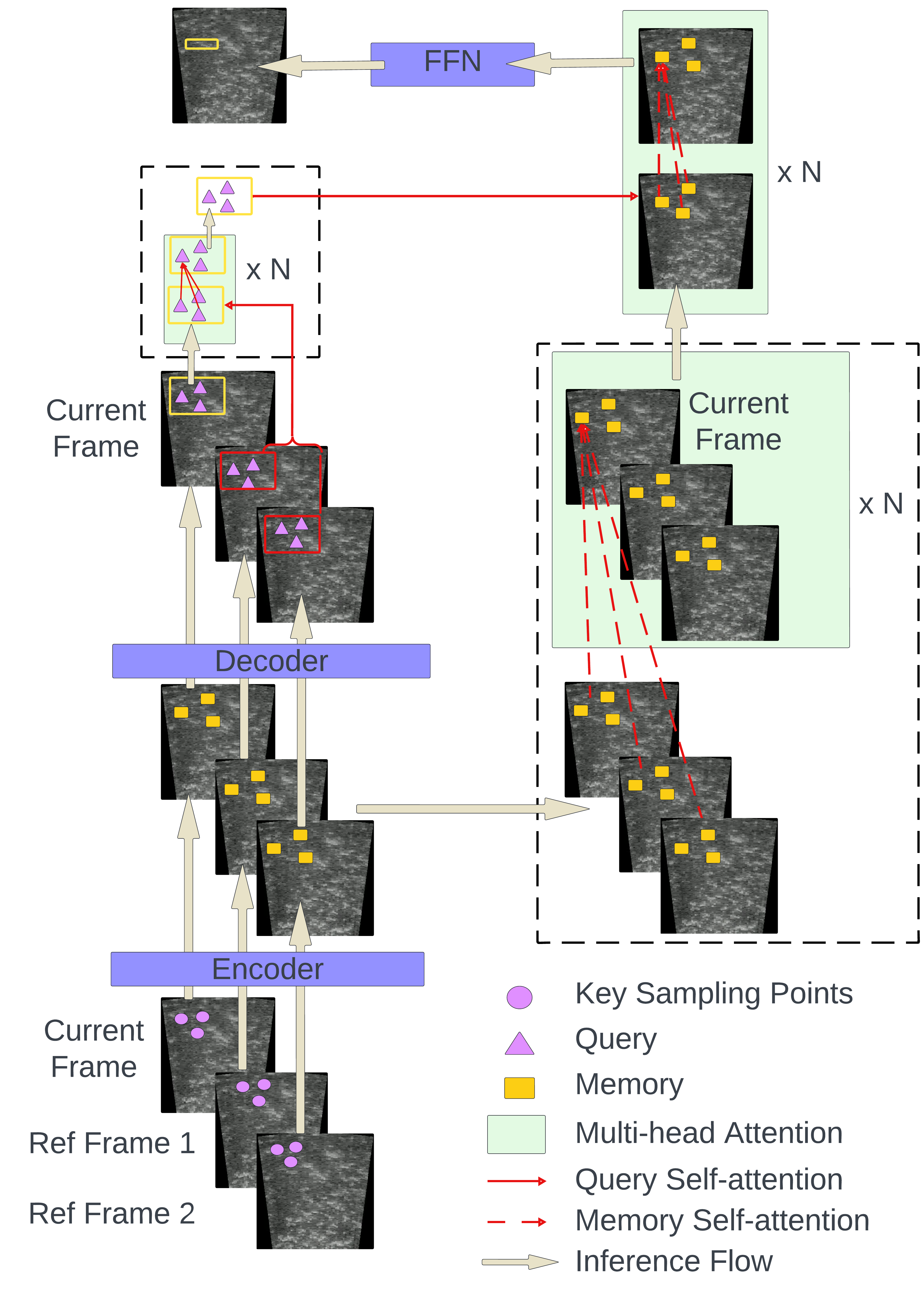}
  \caption{The architecture of Spatial-Temporal Sonar Vision Transformer (STSVT).}
    \vspace{-0.1cm}
  \label{fig:transformer}
  \vspace{-0.3cm}
\end{figure}

\subsection{Cloud Inference with Adapted Spatial-Temporal Transformer}

We then introduce the design of the Spatial-Temporal Sonar Vision Transformer (STSVT) for cloud inference. 

Recent trends in object detection have adopted Transformer architectures to eliminate predefined anchor boxes and many hand-designed components such as non-maximum suppression (NMS), resulting in significant performance improvements \cite{carion2020endtoend, zhu2021deformable, Zhou_2023}. Deformable DETR \cite{zhu2021deformable} uses an attention mechanism to aggregate multi-scale feature maps, enhancing model generalization at different scales. However, our preliminary experiments show that noises, acoustic shadows, and reverberations in sonar frames weaken pixel correlation, negatively impacting the Transformer's point-to-point attention mechanism. 
To address this, incorporating temporal information between frames proves beneficial, as it improves the model’s resilience to noise and allows it to better track moving objects. Accordingly, our adapted Spatial-Temporal Sonar Vision Transformer (STSVT) introduces the following two major components:

\textbf{Spatial Transformer.} We choose the recently proposed Deformable DETR as our still frame detector. To simplify the design, we do not use multi-scale feature representations in either the encoders or decoders of the spatial transformer. Instead, we use only the last stage of the backbone as the input to reduce the complexity. As shown in Figure~\ref{fig:transformer}, the modified detector includes a spatial Transformer encoder and a spatial Transformer decoder, which encodes each frame (including the previous reference frame and current frame) into two compact representations: spatial object query and memory encoding.


\textbf{Temporal Transformer.} The temporal transformer first encodes the spatial details from multiple frames, aggregating this spatial information via a temporal deformable attention mechanism. It then fuses object queries from multiple adjacent frames, selecting relevant object queries and combining them through several self-attention layers. With these intermediate outputs, it then generates detection results. As shown in Figure~\ref{fig:transformer}, we introduce additional encoder-decoder architecture to encode information between frames using memory and object queries.

The primary difference between our Spatial-Temporal Sonar Vision Transformer (STSVT) and existing models lies in the extensive utilization of temporal information between frames. This is particularly crucial for detecting moving objects in sonar data. We set the model to process multiple frames (ranging from 2 to 16) to better capture the displacement information of underwater objects. By doing so, the model can more effectively detect and track moving objects, even in varying underwater conditions.
\begin{figure}
  \includegraphics[width=0.5\textwidth]{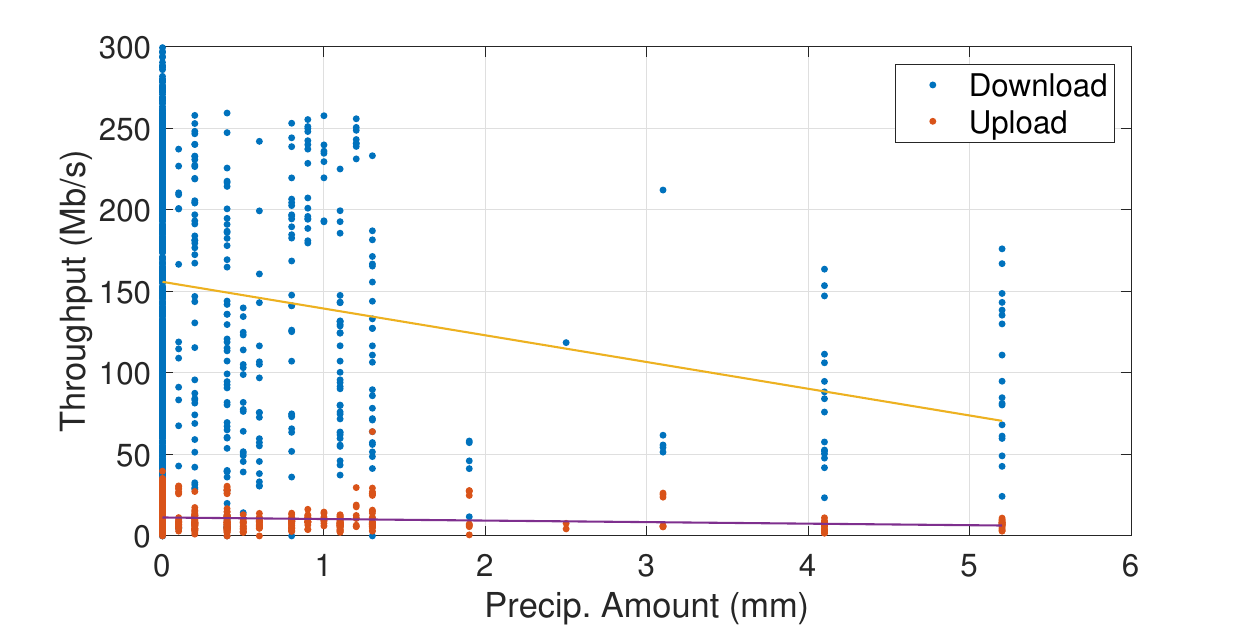}
  \vspace{-0.1cm}
    \captionsetup{width=0.7\linewidth}
  \caption{Throughput and precipitation amount correlations.}
  \label{fig:rain-throughput}
  \vspace{-0.5cm}
\end{figure}

\begin{figure}
  \includegraphics[width=0.30\textwidth]{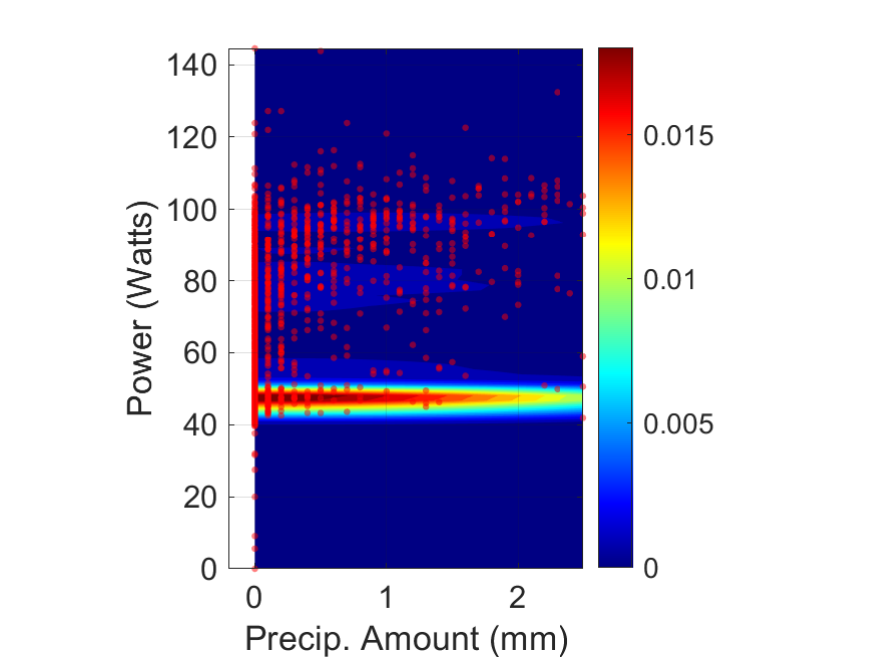}
  \vspace{-0.1cm}
    \captionsetup{width=0.8\linewidth}
  \caption{Satellite power consumption and precipitation amount correlations.}
  \label{fig:rain-power}
    \vspace{-0.5cm}
\end{figure}

\section{Sonar Streaming and Energy Planning}
\label{sec:energy}

When both on-premise and cloud resources are available, cloud inference is preferred for better sonar analytics performance. However, this presents challenges related to data transmission and energy planning. For instance, at one sonar site relying on solar power and satellite communication (e.g., Starlink) for real-time sonar streaming and cloud inference, the lack of a proper energy plan led to a power overload. During peak transmission, the Starlink dish consumed up to 200 Watts, double that of the sonar radar system's 100 Watts, resulting in unexpected contingency costs to repair the power breaker. Additionally, the site uses rechargeable batteries to support overnight operations, but continuous Starlink usage sometimes depleted the battery, causing system outages. These incidents emphasize the need for careful coordination between sonar streaming and energy planning to ensure sustainable operations.

Supporting cloud inference in remote ecosystems, where solar power is often the primary energy source and satellite communication serves as the main network connection, requires careful planning. Adverse weather conditions such as rain, sudden storms, and fluctuating cloud cover can disrupt satellite connectivity and reduce solar power efficiency. To address these challenges, we propose a joint design for sonar streaming and energy planning. The design is based on our observations of how weather conditions impact network connectivity and power availability, and ensure continuous and efficient operations in volatile wild environments.

\subsection{Weather Impact on Satellite Streaming Connectivity}

\begin{figure}
  \includegraphics[width=0.48\textwidth]{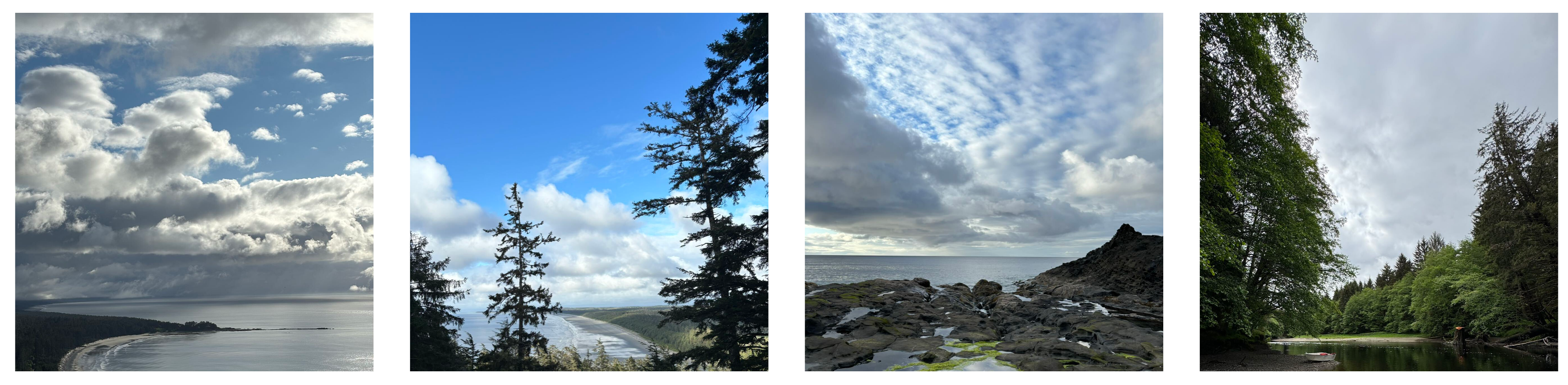}
      \captionsetup{width=0.95\linewidth}
  \caption{Volatile weather patterns such as sudden storms and fluctuating cloud cover. From left to right: cumulus, altocumulus, cirrocumulus, and nimbostratus.}
  \label{fig:cloud}
  \vspace{-0.3cm}
\end{figure}

\begin{figure}
  \includegraphics[width=0.40\textwidth]{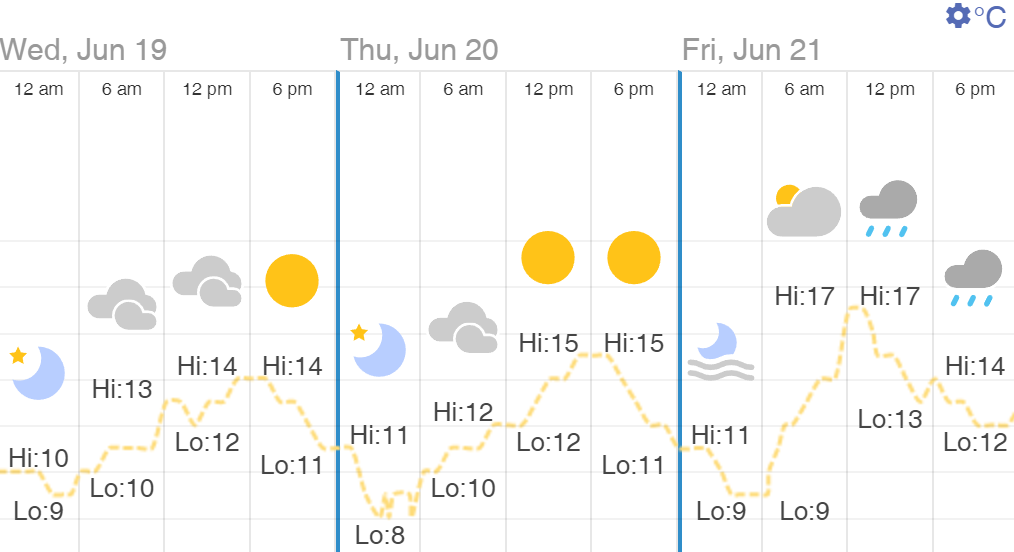}
        \captionsetup{width=1\linewidth}
  \caption{Volatile weather patterns, 4-hour window forecast. We use a more fine-grained image-based forecast method.}
  \label{fig:forecast}
  \vspace{-0.3cm}
\end{figure}
We have observed that Starlink’s throughput can be significantly affected by weather conditions. As shown in Figure~\ref{fig:rain-throughput}, our results indicate an inverse correlation between throughput and precipitation amount. Specifically, throughput drops by an average of 15\% during any form of precipitation. Throughput is particularly constrained during heavy rain events (> 4 mm per hour), affecting both downloads and uploads. The primary cause of this decrease in throughput is rain attenuation. Ka- and Ku-band radio waves, which are used by Starlink, are particularly susceptible to degradation in the presence of rain. The attenuation of these radio waves leads to a significant reduction in the signal quality received by the satellite terminals. Additionally, the presence of clouds can further interrupt data communications. Even light clouds can reduce satellite signal strength by approximately 10\%, with thicker clouds associated with heavy rainfall further obstructing the network paths in both uplinks and ground-satellite links.

We also find that when streaming sonar data, the current dish’s power consumption averages around 51.3 Watts but can go as high as 166.5 Watts, with the dish’s power consumption positively correlated with precipitation. We present the density analysis in Figure~\ref{fig:rain-power}. It is clear that, without rain, power consumption is lower in general, with occasional spikes due to other factors such as tuning the dish direction or establishing connections with farther away satellites. When there is rain or heavy clouds, power consumption becomes persistently higher, likely due to the interference from the rain or clouds.

\subsection{Sustainable Energy Planning}
In parallel, solar PV energy production is also highly sensitive to weather conditions. Solar panels generate electricity based on the amount of sunlight they receive, which can be drastically reduced by cloud cover, rain, or snow. Figures~\ref{fig:cloud} and~\ref{fig:forecast} show the volatile weather patterns on-site,  highlighting the sudden changes and fluctuating cloud cover. The variability in solar energy production necessitates the need for efficient energy management strategies to ensure a stable power supply and limit peak energy use, especially in off-grid setups in wild ecosystems.

\begin{figure}
  \centering
  \begin{subfigure}[b]{0.22\linewidth}
    \includegraphics[width=\linewidth]{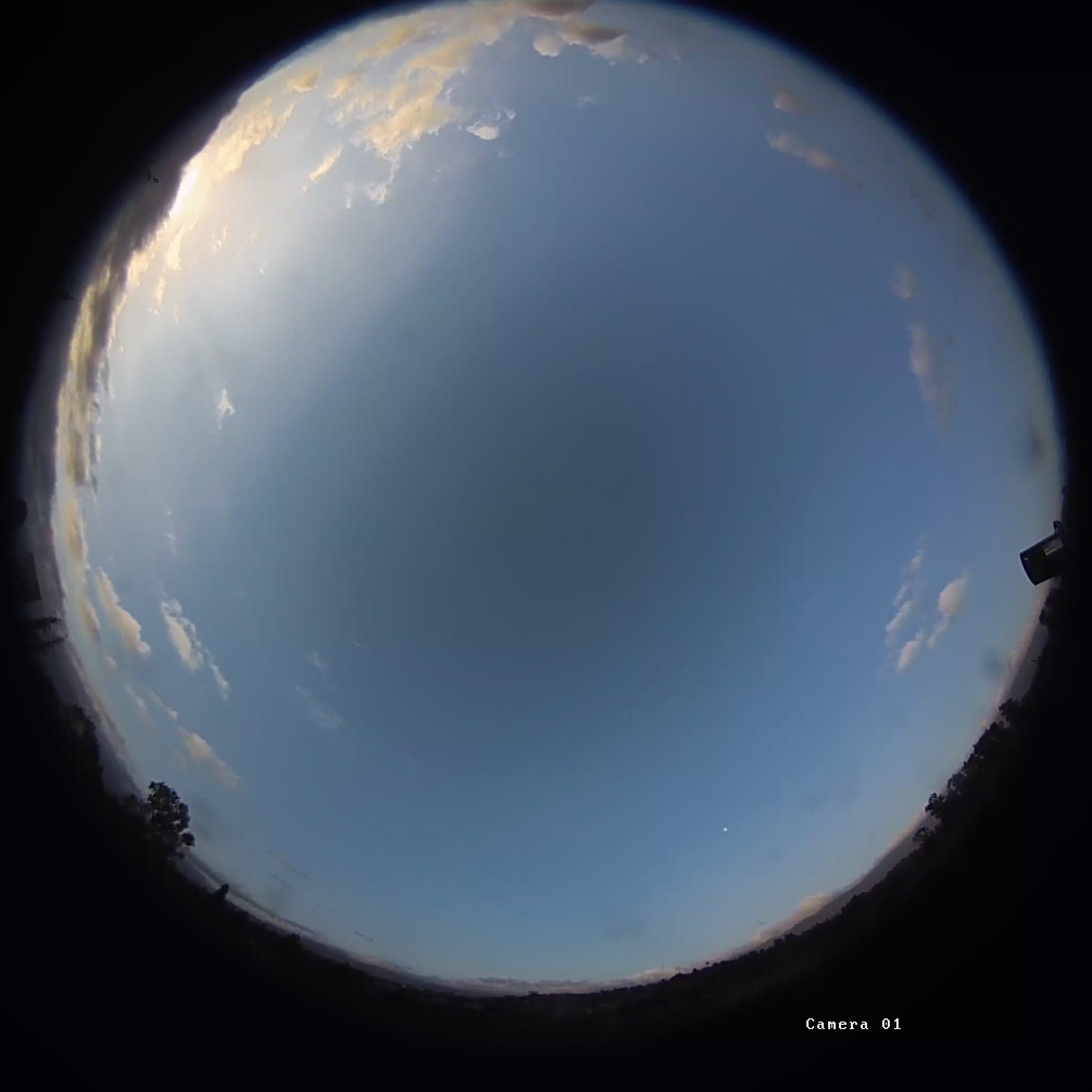}
    \caption{}
    \label{fig:lens-clear}
  \end{subfigure}
  \hspace{0.01\linewidth}
  \begin{subfigure}[b]{0.22\linewidth}
    \includegraphics[width=\linewidth]{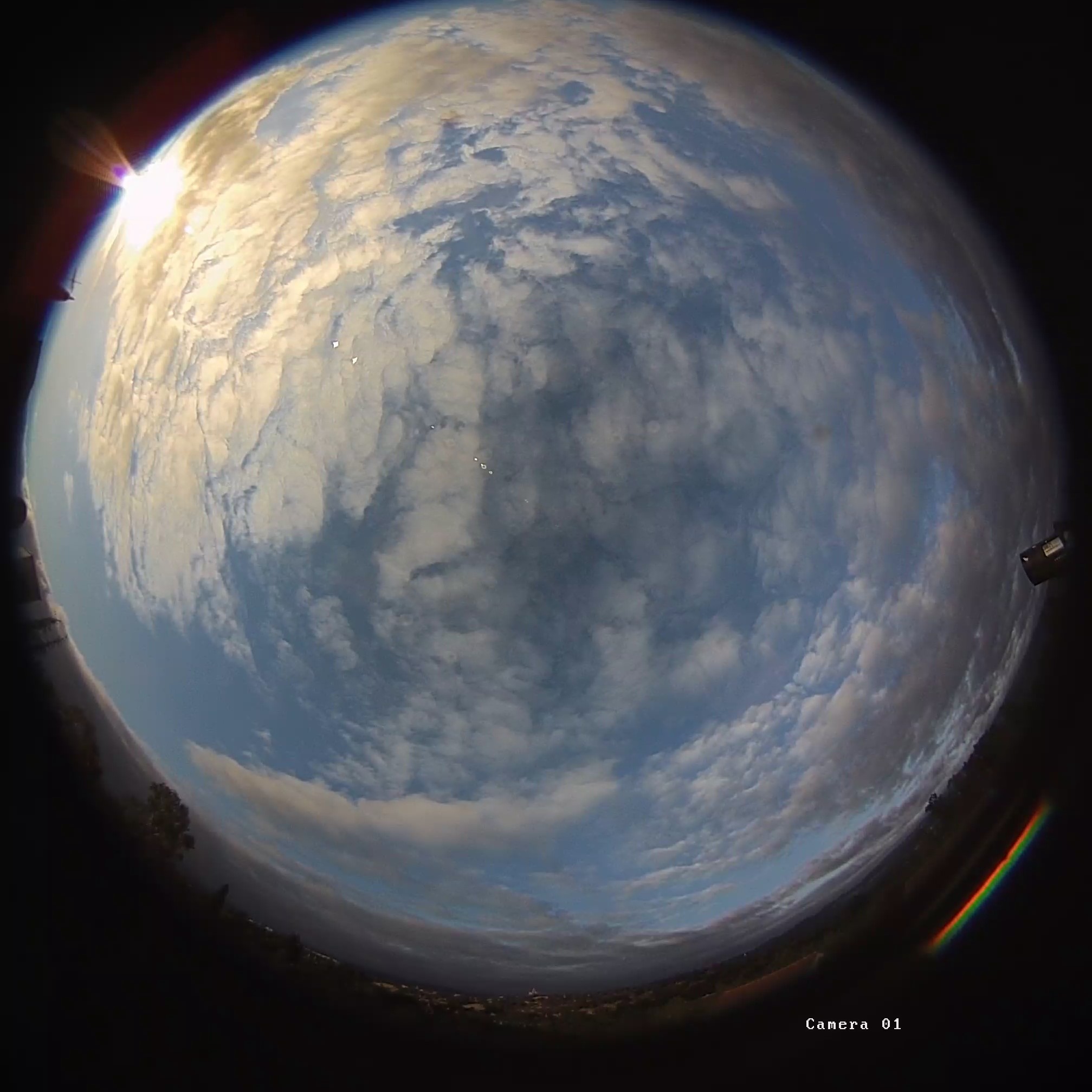}
    \caption{}
    \label{fig:lens-sparse}
  \end{subfigure}
    \hspace{0.01\linewidth}
  \begin{subfigure}[b]{0.22\linewidth}
    \includegraphics[width=\linewidth]{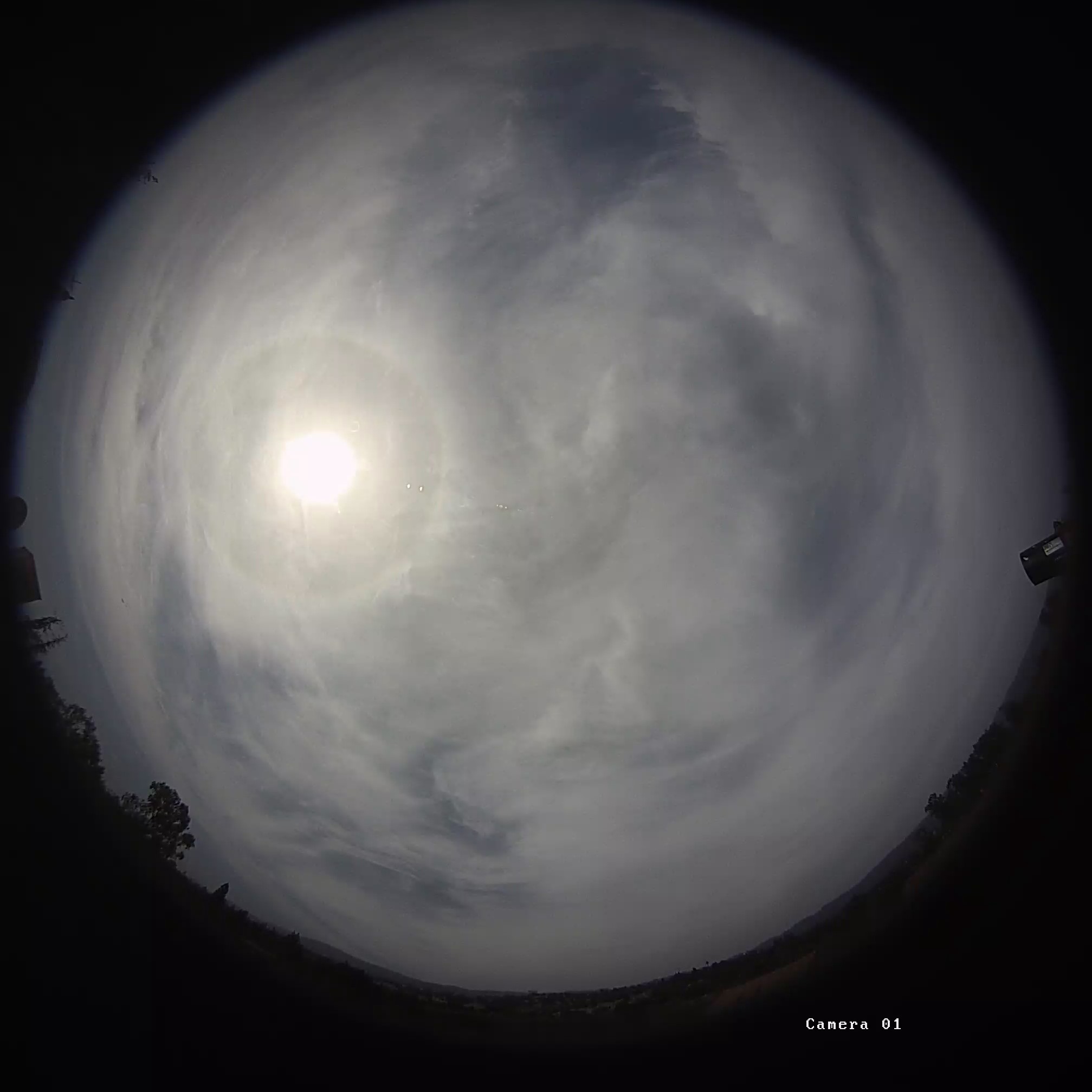}
    \caption{}
    \label{fig:lens-thick}
  \end{subfigure}
  \hspace{0.01\linewidth}
  \begin{subfigure}[b]{0.22\linewidth}
    \includegraphics[width=\linewidth]{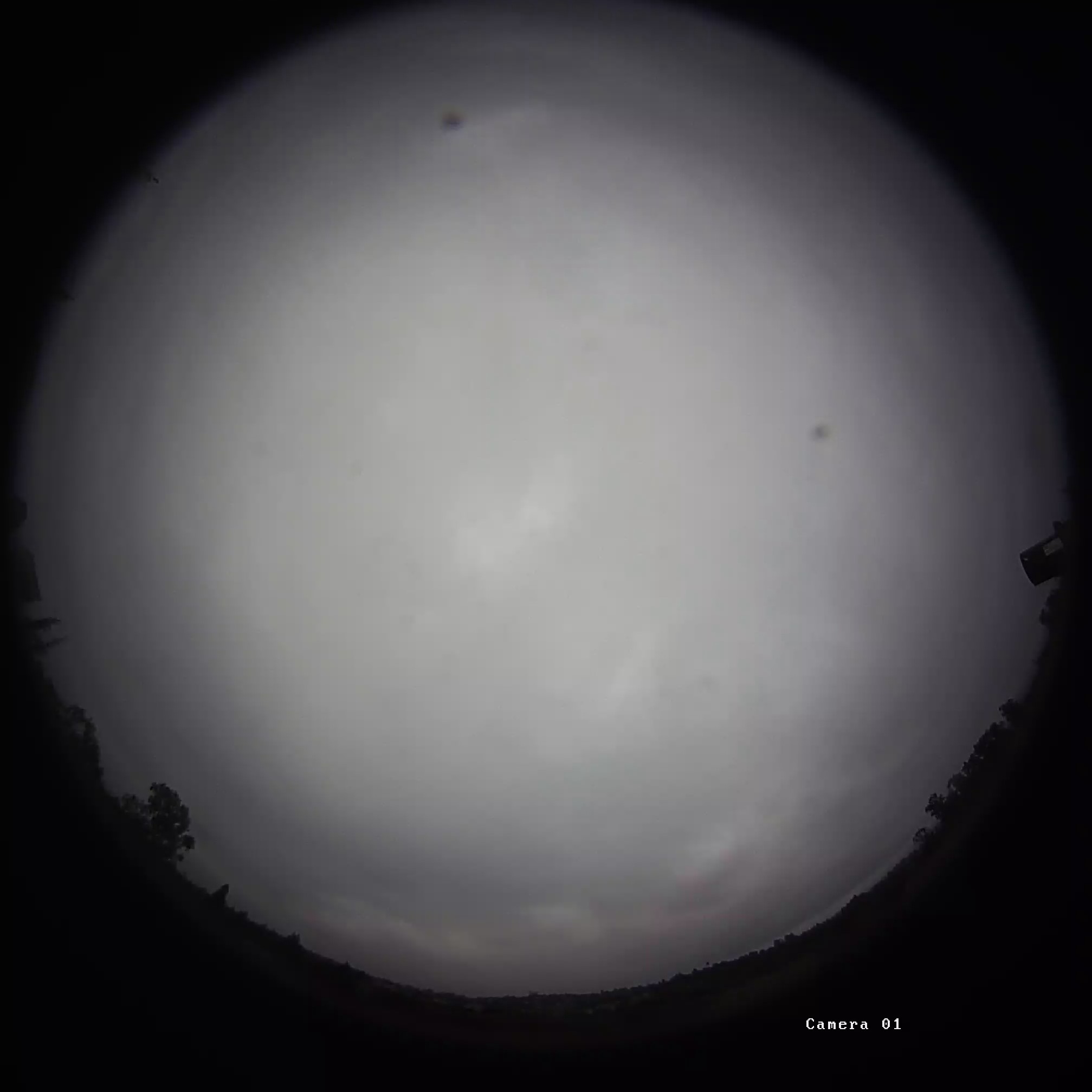}
    \caption{}
    \label{fig:lens-full}
  \end{subfigure}
  \caption{Fisheye lens monitoring of volatile weather patterns: \textbf{(a)} clear sky, \textbf{(b)} sparse, \textbf{(c)} thick, \textbf{(d)} full.}
  \label{fig:lens}
  \vspace{-0.5cm}
\end{figure}

We utilize a short-term PV output forecast model~\cite{nie2020pv} that learns a mapping from the sky image to future PV power output. This mapping is trained on fisheye lens images. Several examples are shown in Figure~\ref{fig:lens}. These sky images are frames captured by a 6-megapixel 360-degree fish-eye Hikvision DS-2CD6362F-IV27 camera. This network camera is a cost-effective, low-power option (15 Watts) for sky monitoring. It captures at 2048 × 2048 pixels resolution at 20 fps. The PV output generation data are logged with a 1-minute frequency and are minutely averaged. To further optimize performance, we fine-tune the forecast model using our captured sky images and power data from three 300W PV modules with 18.6\% efficiency, a typical configuration for powering sonar sites in wild ecosystems.

\subsection{Multi-Stratum Streaming Optimization}

Sonar frames are usually rectangular with a high height-width ratio, which poses a challenge for efficient streaming and processing. In this work, we explore a multi-stratum streaming mechanism by splitting the rectangle into multiple near-square strata, each streamed with different configurations. This approach improves streaming efficiency and detection performance. Another compelling argument for performing multi-stratum streaming is the significant impact of frame aspect ratio on inference results. Using images with an aspect ratio close to a square, rather than a significant width-height difference, leads to better detection performance. When images are resized to fixed sizes (e.g., 416x416 or 640x640) as inputs, large aspect ratio differences can cause distortion, negatively affecting the detection models' ability to learn object features accurately. Square-like images minimize this distortion, maintaining the target objects' original proportions and improving detection accuracy, which is crucial since many objects in sonar frames are thin and long. Additionally, CNN models process images at a uniform scale, so near-square images ensure an even distribution of computational resources, enhancing efficiency and effectiveness. 
\begin{figure}

  \includegraphics[width=0.44\textwidth]{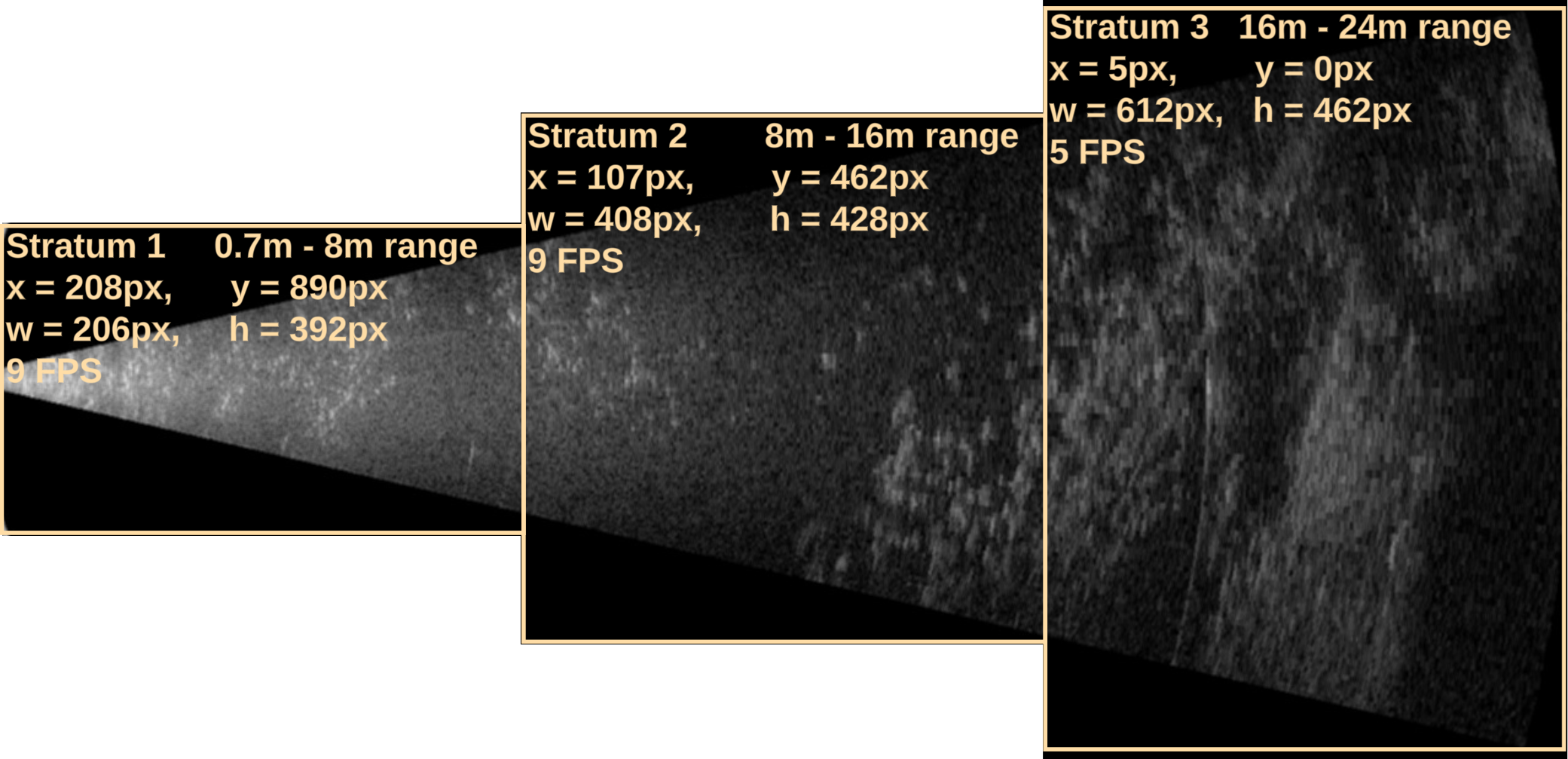}
     \captionsetup{width=1\linewidth}
  \caption{An example of a 3-stratum split in sonar frame.}
  \label{fig:stratum}
  \vspace{-0.5cm}
\end{figure}




As shown in Figure~\ref{fig:stratum}, we propose a multi-stratum streaming framework that integrates satellite connectivity metrics and PV energy production forecasts to adjust streaming configurations, such as downscaling factors, framerates, and preprocessing filters, all of which impact final data rates. This approach allows for dynamic adjustment of both internet usage and energy consumption based on current and forecast conditions. 


We formulate the multi-stratum streaming optimization problem as follows: consider a sonar stream that is segmented into \( N \) strata. Each stratum \( i \in [1, \ldots, N] \) introduces a set of control parameters \( s_{i,j} \). A configuration consisting of \( M \) selections for stratum \( i \) is \( c = [s_{1,1}, s_{1,2}, s_{1,M}, \ldots, s_{i,M}] \), representing a specific combination of these control parameters. The set of all configurations is denoted by \( \mathcal{S} \). For any given configuration \( c \), we define three critical metrics: the bandwidth usage \( B(c) \), the analytics performance \( A(c) \), and the power consumption \( P(c) \). The goal of profiling is to identify configurations that are \textit{Pareto-optimal}. A configuration \( c \) is considered Pareto-optimal if there is no other configuration \( c' \) that simultaneously uses less bandwidth, less power, and achieves higher analytics performance. Formally, the set of Pareto-optimal configurations \( \mathcal{P} \) is defined as follows:

\begin{equation}
\begin{aligned}
\mathcal{P} = \left\{ c \in \mathcal{S} : \nexists~c' \in \mathcal{S} \ \text{such that} \right. & B(c') < B(c), \ P(c') < P(c), \\ \text{and} & \left. A(c') > A(c) \right\}
\end{aligned}
\end{equation}

To solve the optimization problem, we first identify the key configurations that impact sonar analytics, such as frame size scaling factor, frame rate, and power-related operations. By focusing on these parameters, we can limit the search space for efficient computing. To achieve this, we discretize the parameter space and evaluate all possible combinations. Specifically, we employ an exhaustive search over the discrete parameter space to identify Pareto-optimal configurations. However, if the parameter space were to be treated as continuous, more advanced search techniques~\cite{goldberg1989genetic, kirkpatrick1983optimization} would be necessary to efficiently navigate the vast search space. 

\subsection{Continuous Operation}
In our streaming optimization, the bandwidth constraint is determined through continuous monitoring of the Starlink connection, ensuring that the selected configurations do not exceed the available bandwidth. Similarly, the power consumption constraint considers both current energy production and future forecasts. These forecasts are used to estimate energy availability over upcoming periods, ensuring that the chosen configuration remains sustainable over the forecasted timeframe. This approach effectively prevents unexpected shutdowns or system outages due to energy depletion. It is noted that in the current system design, we have accounted for sufficient safety margins for rechargeable battery power to support overnight operations. Specifically, energy planning is conducted during the day when energy is more abundant, and configurations are aware of reserving additional energy during adverse weather conditions or low energy production. This ensures that the system maintains continuous and stable operations throughout the night. Thus, even under constrained energy conditions, our system remains reliable during the overnight period.

\begin{table}[t]
\centering
\renewcommand{\arraystretch}{1.1}
\captionsetup{justification=centering}
\caption{YOLOv8x with different inputs.}
\label{table:apr_yolo}
\footnotesize
\begin{tabular}{lccc} 
\toprule
\textbf{Metrics}   & \textbf{Raw Channel} & \textbf{CFC Channels} & \textbf{Our Channels} \\
\midrule
AP IoU=0.50        & 0.834 & 0.862 & \textbf{0.876 (+0.014)}\\
AP IoU=0.50:0.95   & 0.407 & 0.423 & \textbf{0.451 (+0.023)}\\
Precision          & 0.895 & 0.901 & \textbf{0.909 (+0.008)}\\
Recall             & 0.766 & 0.788 & \textbf{0.805 (+0.017)}\\
val/box\_loss      & 1.631 & 1.598 & \textbf{1.550 (-0.048)}\\
val/class\_loss    & 1.049 & 0.976 & \textbf{0.892 (-0.084)}\\
\bottomrule
\end{tabular}
\end{table}

\begin{table}[t]
\centering
\renewcommand{\arraystretch}{1.1}
\captionsetup{justification=centering}
\caption{STSVT with different inputs. }
\label{table:apr_stsvt}
\footnotesize
\begin{tabular}{lccc|ccc} 
\toprule
\textbf{} & \textbf{IoU} & \textbf{Area} & \textbf{MaxDet} & \textbf{Raw} & \textbf{CFC} & \textbf{Ours} \\
\midrule
AP & 0.50      & all & 100 & 0.507 & 0.574  & \textbf{0.591 (+0.017)} \\
AP & 0.50:0.95 & all & 100 & 0.155 & 0.189 & \textbf{0.219 (+0.030)} \\
AP & 0.75      & all & 100 & 0.048 & 0.055 & \textbf{0.107 (+0.052)} \\
AP & 0.50:0.95 & small & 100 & 0.075 & 0.089 & \textbf{0.110 (+0.021)}\\
AP & 0.50:0.95 & medium & 100 & 0.201 & 0.242 & \textbf{0.444 (+0.202)}\\
AP & 0.50:0.95 & large & 100 & 0.314 & 0.424 & \textbf{0.826 (+0.402)}\\
\midrule
AR & 0.50:0.95    & all & 1 & 0.133 & 0.155 & \textbf{0.186 (+0.031)}\\
AR & 0.50:0.95    & all & 10 & 0.233 & 0.278 & \textbf{0.327 (+0.049)}\\
AR & 0.50:0.95    & all & 100 & 0.267 & 0.314 & \textbf{0.357 (+0.043)}\\
AR & 0.50:0.95    & small & 100 & 0.146 & 0.204 & \textbf{0.210 (+0.006)}\\
AR & 0.50:0.95    & medium & 100 & 0.338 & 0.379 & \textbf{0.444 (+0.065)}\\
AR & 0.50:0.95    & large & 100 & 0.328 & 0.420 & \textbf{0.880 (+0.460)}\\
\bottomrule
\end{tabular}
\vspace{-0.3cm}
\end{table}

\section{System Evaluations}
\label{sec:eval}
\subsection{Implementation and Configuration}
We collaborated closely with biologists, forest technologists, and electricians to set up the system for continuous monitoring and surveillance at two different rivers in British Columbia, Canada: YK River and KN River. At YK River, we used the ARIS Explorer 1800 Sonar, which has an effective detection range of 35 meters. While at KN River, we employed the ARIS Explorer 3000 Sonar, with a detection range of 15 meters. The edge device used was a Jetson ORIN Nano 8GB. Onsite solar power was provided by three 300W 18.6\% efficiency Q. Peak-G4.1 PV modules. The cloud server setup included a Lambda Vector Server equipped with four A5000 GPUs. The system was deployed for six months, monitoring underwater environments. As noted in Section 3.2, we also employed a third-party annotation service to label multiple object-tracking annotations for all objects in the converted grayscale sonar clips, including salmon, otter, and smolt. These annotations were used for training and fine-tuning the models, ensuring reliable performance.


\subsection{Channel Population Impacts}

We used our channel population results for training and validating on the YK dataset, comparing sonar raw channel, CFC channels (~\cite{kay2022caltech}, considered as SOTA), and our channels. As shown in Table~\ref{table:apr_yolo}, our channels consistently outperformed both the raw and CFC channels across all metrics. Average Precision (AP) at IoU 0.50 improved from 0.834 (raw) and 0.862 (CFC) to 0.876 with our channels. AP at IoU 0.50:0.95 increased from 0.407 (raw) and 0.423 (CFC) to 0.451. Validation losses for bounding boxes and classification were also reduced significantly with our channels. All these results demonstrate the superior performance of our channel population method, highlighting its efficacy in enhancing sonar data detection.

\begin{figure}
  \includegraphics[width=0.40\textwidth]{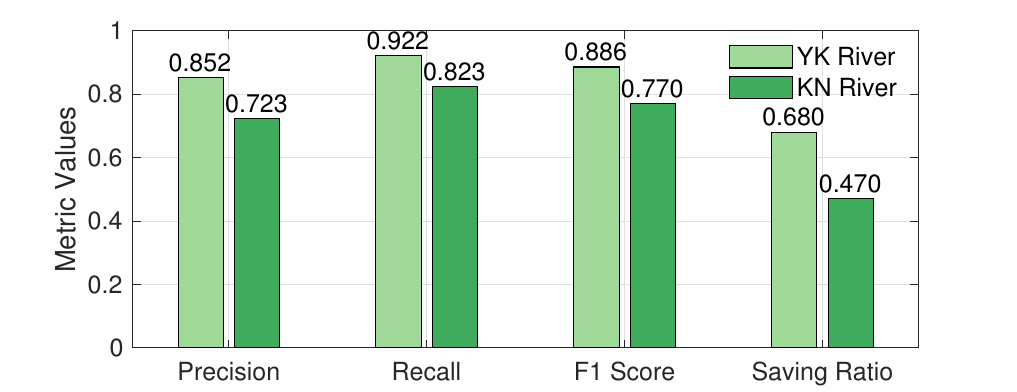}
  \caption{Motion detection results.}
  \label{fig:motion}
  \vspace{-0.3cm}
\end{figure}

\begin{table}[t]
\centering
\renewcommand{\arraystretch}{1.1}
\captionsetup{justification=centering}
\caption{Comparison of CFC and STSVT.}
\label{table:cfc_stsvt}
\footnotesize
\begin{tabular}{lccc} 
\toprule
\textbf{Metrics}  & \textbf{AP IoU=0.50} & \textbf{AP IoU=0.50:0.95} & \textbf{val/box\_loss} \\
\midrule
CFC & 0.640 & 0.271 & 2.173\\
STSVT & \textbf{0.735 (+0.095)} & \textbf{0.334 (+0.063)} & \textbf{0.251 (-1.922)}\\
\bottomrule
\end{tabular}
\end{table}

Similarly, the experiment results presented in Table \ref{table:apr_stsvt} show a comprehensive comparison of average precision (AP) and average recall (AR) metrics with different inputs: Raw, CFC, and Ours. This experiment was conducted using Spatial-Temporal Sonar Vision Transformer (STSVT) as the detector on the KN river dataset, which is more challenging. Note in this figure, "small" refers to objects smaller than $32 \times 32$ pixels, "medium" refers to objects between $32 \times 32$ and $72 \times 72$ pixels, and "large" refers to objects larger than $72 \times 72$ pixels. "MaxDet" refers to the maximum number of detections considered per frame during evaluation.

\begin{table*}
  \caption{On-premise and cloud inference comparison.}
  \label{tab:edgecloud}
  \footnotesize
  \begin{tabular}{ccccccccccc}
    \toprule
    Stratum & Model & Format & Param & Where & GPU & AP IoU=0.50 & val/box\_loss & Edge Power & Transmission Power & Power Sum\\
    \midrule
    \multirow{2}{*}{1} & YOLOv8m & TensorRT & 25.84M & Edge & Ampere & 0.379 & 1.581 & 9.34W & N/A & 9.34W\\
    & STSVT & default & 53.47M & Cloud & A5000 & 0.507 & 0.202 & 5.35W & 45.48W & 50.83W\\
    \midrule
    \multirow{2}{*}{2} & YOLOv8m & TensorRT & 25.84M & Edge & Ampere & 0.414 & 1.314 & 9.68W & N/A & 9.68W\\
    & STSVT & default & 53.47M & Cloud & A5000 & 0.574 & 0.169 & 5.74W & 47.55W & 53.39W\\
    \bottomrule
  \end{tabular}
\end{table*}

On this dataset, our YOLOv8x model running on edge only achieves below 0.4 in AP IoU=0.5. When using STSVT, our channel inputs consistently outperform both the raw and CFC channels across all evaluated metrics. Specifically, when examining AR metrics, our method again demonstrates superior performance. For instance, AR at IoU=0.50:0.95 for all areas with MaxDets=100 is 0.357 with our method, compared to 0.267 for the raw channel and 0.314 for the CFC channels. Notably, our method's AR for large objects (> 72px $\times$ 72px) with MaxDets=100 reaches an impressive 0.880. Since most real-world objects of interest in sonar frames are larger than this threshold, this highlights our method's exceptional ability to detect and recall.

We present motion detection results based on Canny edge detectors over our populated channel, comparing Precision, Recall, and F1 Score in Figure~\ref{fig:motion}. To ensure no objects were missed, we tuned the Canny parameter to prioritize high Recall. The results on the YK river and KN river datasets also show the saving ratio, indicating the proportion of non-motion frames that do not need processing or uploading. This helps save processing resources and reduces network transmission requirements.



\subsection{Detection Performance}

Table~\ref{table:cfc_stsvt} compares the performance of CFC and STSVT methods on the KN river dataset using AP and validation loss metrics. STSVT achieves an AP of 0.735 at IoU=0.50, showing a significant 0.095 improvement over CFC's 0.640. For AP at IoU=0.50:0.95, STSVT outperforms CFC by 0.063, reaching 0.334. Additionally, STSVT significantly reduces validation bounding box loss to 0.251, a decrease of 1.922 from CFC's 2.173, indicating improved prediction accuracy and model effectiveness across different IoU thresholds.

Table~\ref{tab:edgecloud} provides a comparison of on-premise edge inference and cloud inference across different configurations. When a single frame is considered as one stratum, edge inference with YOLOv8m shows lower power consumption (9.34W) but lower precision compared to cloud inference with STSVT, which achieves a higher AP at IoU=0.50 (0.507) at a significantly higher power cost (50.83W) due to additional transmission power over Starlink. In the scenario where the frame is split into two square strata, YOLOv8m at the edge again consumes less power (9.68W) compared to cloud inference with STSVT (53.39W), which shows improved precision but at the cost of increased power consumption due to cloud transmission. This demonstrates that while cloud inference enhances model performance, edge inference remains more energy-efficient for scenarios where power consumption is a critical factor.

The results in Table \ref{tab:freq} compare the cloud models and edge models in terms of inference time. For edge deployment, the YOLOv8m model in TensorRT format offers the fastest inference at 65.3 ms, supporting real-time processing at around 15 fps. In contrast, cloud deployment using the A5000 GPU drastically reduces inference times, with the YOLOv8m and YOLOv8x models achieving 8.59 ms and 9.17 ms, respectively, suitable for frame rates up to 60 fps. The STSVT model, although slower at 56.1 ms, still supports real-time processing at 15 fps. These speeds confirm that both edge and cloud deployments of SALINA can handle sonar analytics in real time, with cloud solutions offering higher frame rate capabilities.
\subsection{Tracking Performance}

\begin{table}

  \caption{Inference time comparison. }

  \label{tab:freq}
  \footnotesize
  \begin{tabular}{cccccc}
    \toprule
    Model & Format & Param & Where & GPU & Inference Time \\
    \midrule
    YOLOv8m & default & 25.84M & Edge &  Ampere & 81.7 $\pm$ 8.15ms\\
    YOLOv8m & ONNX &  25.84M & Edge & Ampere & 78.6 $\pm$ 7.30ms\\
    YOLOv8m & TensorRT & 25.84M & Edge & Ampere & 65.3 $\pm$ 5.72ms\\
    \midrule
    YOLOv8x & default & 68.2M & Cloud & A5000 & 9.17 $\pm$ 1.13ms\\
    YOLOv8m & default & 25.84M & Cloud & A5000 & 8.59 $\pm$ 0.92ms \\
    STSVT & default & 53.47M & Cloud & A5000 & 56.1 $\pm$ 10.0ms \\
  \bottomrule
\end{tabular}
\end{table}

\begin{figure}
  \includegraphics[width=0.35\textwidth]{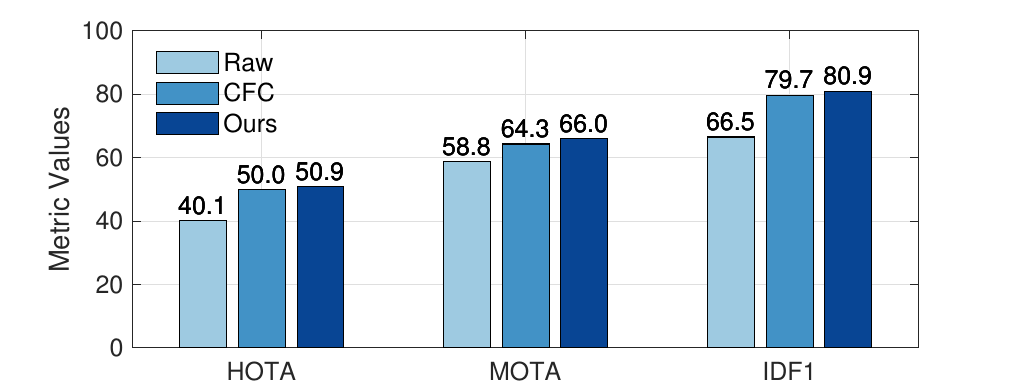}
  \caption{HOTA, MOTA, and IDF1 comparison.}
  \label{fig:tracking}
  \vspace{-0.5cm}
\end{figure}

We further evaluate multi-object tracking performance, where our channel inputs demonstrate superior results compared to both the raw and CFC channels. The experiments were conducted on the KN river dataset. 

The tracking metrics compared in our experiments are HOTA~\cite{luiten2021hota}, MOTA~\cite{bernardin2008evaluating}, and IDF1~\cite{ristani2016performance}:
\begin{itemize}
    \item HOTA (Higher Order Tracking Accuracy) evaluates the balance between object detection and association performance, incorporating metrics such as detection accuracy, association accuracy, and localization accuracy. 
    \item MOTA (Multiple Object Tracking Accuracy) measures overall tracking performance by accounting for false positives, false negatives, and identity switches. It emphasizes detection quality and is widely used in tracking benchmarks.
    \item IDF1 evaluates the accuracy of maintaining consistent object identities over time. 
\end{itemize}

\begin{figure*}
  \centering
  \begin{subfigure}[b]{0.30\linewidth}
    \includegraphics[width=\linewidth]{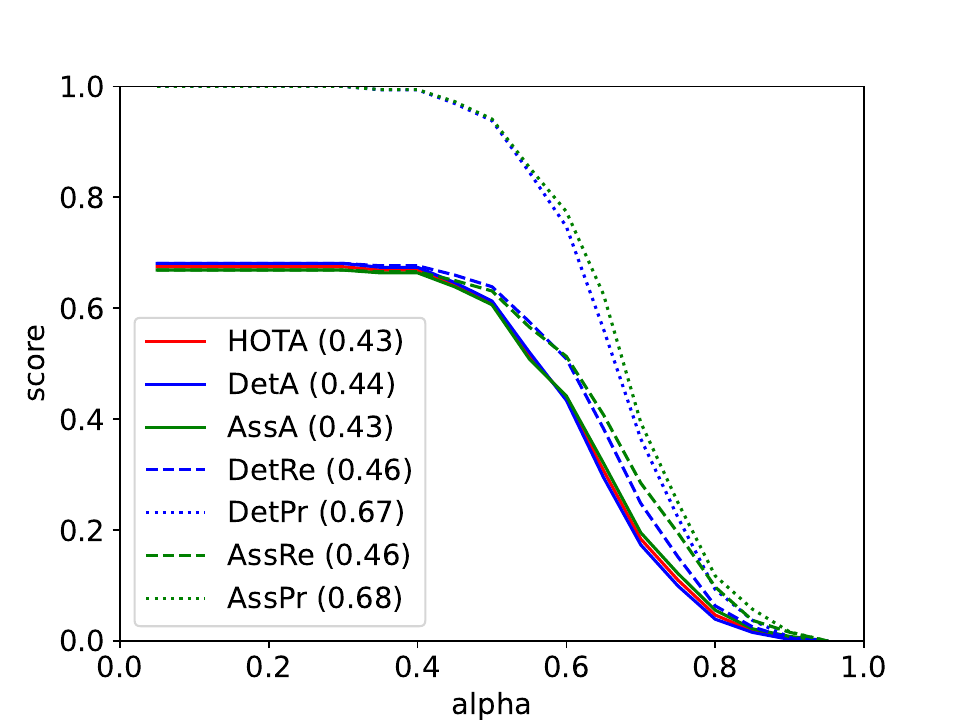}
    \caption{Raw}
    \label{fig:mtrack_raw}
  \end{subfigure}
  \hspace{0.01\linewidth}
  \begin{subfigure}[b]{0.30\linewidth}
    \includegraphics[width=\linewidth]{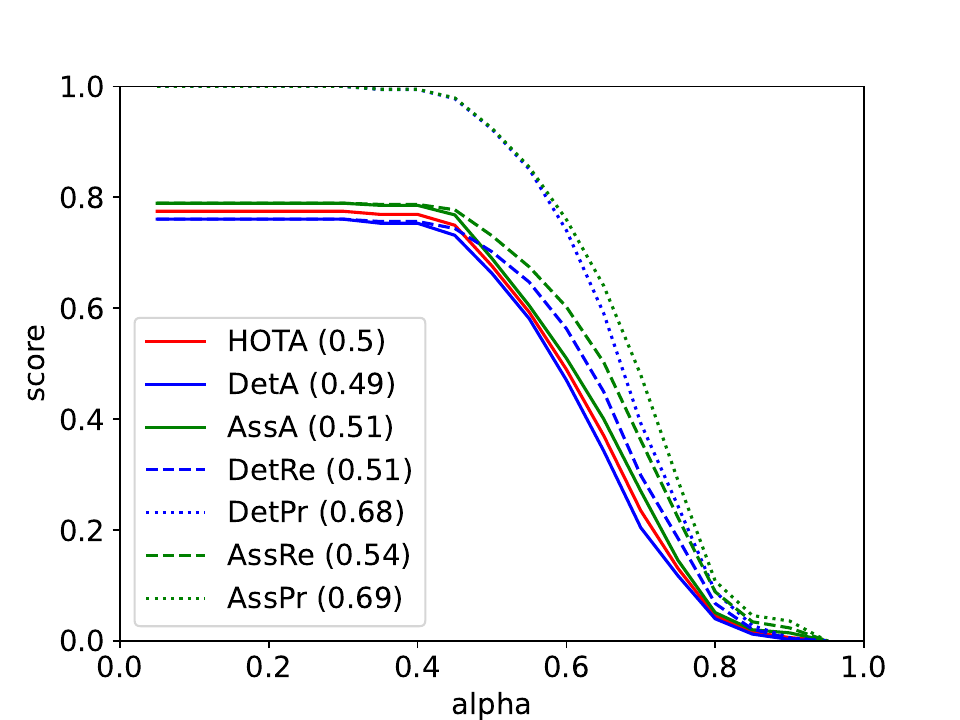}
    \caption{CFC}
    \label{fig:mtrack_cfc}
  \end{subfigure}
    \hspace{0.01\linewidth}
  \begin{subfigure}[b]{0.30\linewidth}
    \includegraphics[width=\linewidth]{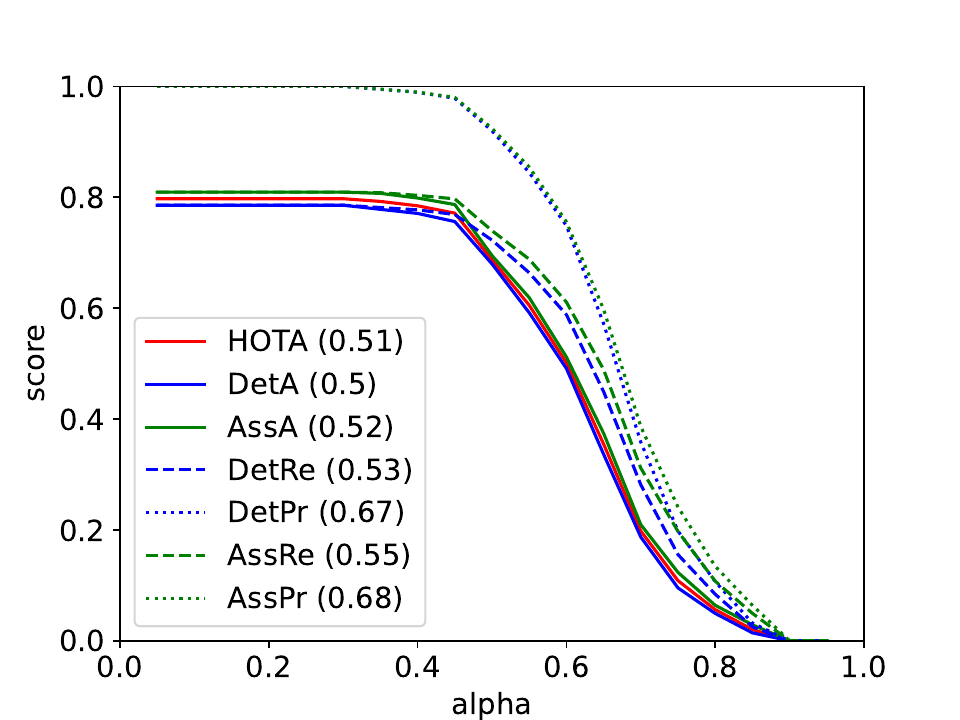}
    \caption{Ours}
    \label{fig:mtrack_our}
  \end{subfigure}
  \vspace{-0.1cm}
  \caption{Tracking Performance Breakdown.}
  \label{fig:mtrack}
  \vspace{-0.3cm}
\end{figure*}

As shown in Figure~\ref{fig:tracking}, we achieved a HOTA score of 50.868, indicating higher accuracy in tracking objects over time, compared to 49.983 for CFC and 40.072 for the raw channel. In terms of MOTA, our channel inputs also excel, with a score of 65.966, surpassing CFC's 64.286 and the raw channel's 58.824. Additionally, our channel inputs maintain object identities exceptionally well, as evidenced by an IDF1 score of 80.941, which is higher than CFC's 79.714 and significantly better than the raw channel's 66.5. These results, with a relative increase of 10.1\%, confirm the effectiveness of our channel inputs in delivering more accurate and reliable multi-object tracking performance.

We present a detailed tracking performance breakdown in Figure~\ref{fig:mtrack}, showing six components of the HOTA metric: DetA, measuring single-frame detection accuracy; AssA for cross-frame association accuracy; DetRe and DetPr for detection recall and precision; and AssRe and AssPr for association recall and precision. We varied the alpha parameter in HOTA to adjust the emphasis between detection and association, illustrating how changes in alpha impact each metric. The evaluation results demonstrate that our channel input outperforms both the Raw channel and state-of-the-art CFC channels, with improvements across detection and association metrics, highlighting the efficacy of our novel channel population pipeline.


\subsection{Sustainable Streaming Performance}
\begin{figure}
  \centering
  \begin{subfigure}[b]{1\linewidth}
    \includegraphics[width=\linewidth]{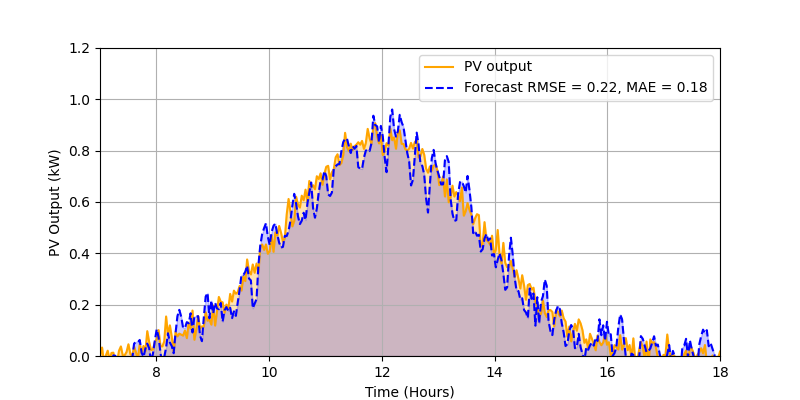}
    \caption{Sunny Day}
    \label{fig:sunny1}
  \end{subfigure}
  \begin{subfigure}[b]{1\linewidth}
    \includegraphics[width=\linewidth]{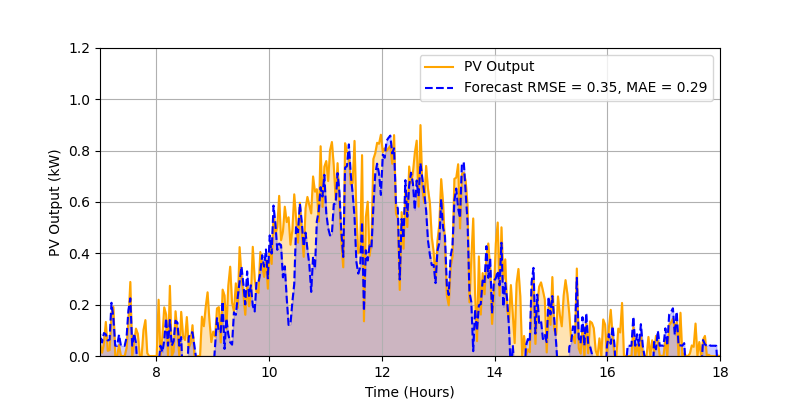}
    \caption{Cloudy Day}
    \label{fig:cloudy1}
  \end{subfigure}
  \vspace{-0.1cm}
  \caption{Short-term solar PV power forecast.}
  \label{fig:pv}
  \vspace{-0.3cm}
\end{figure}

\begin{figure}
  \includegraphics[width=0.40\textwidth]{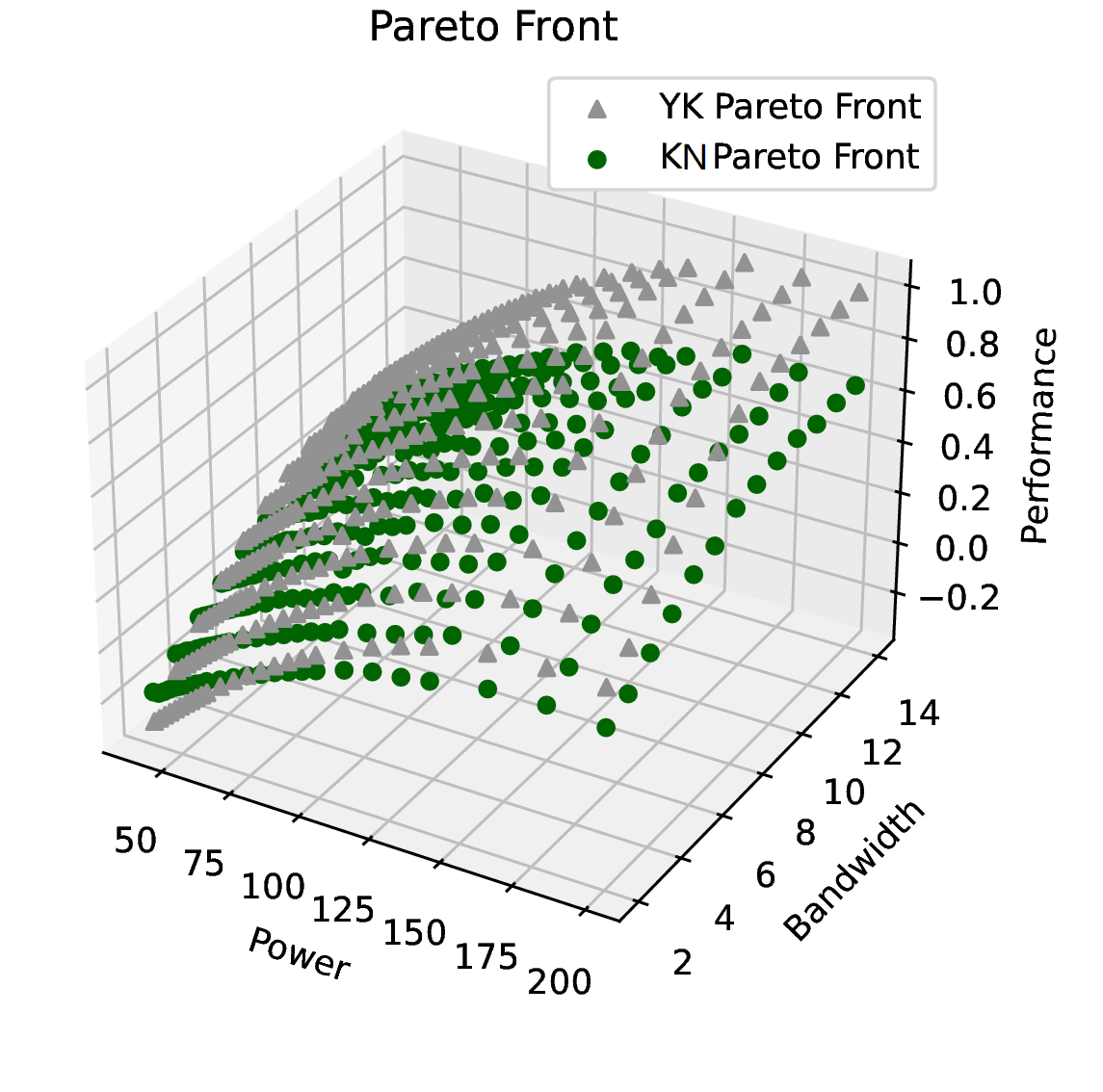}
  \caption{Pareto front at YK River and KN River.}
  \label{fig:pareto}
  \vspace{-0.5cm}
\end{figure}

In Figure~\ref{fig:pv}, we show short-term solar PV power forecasts on one day. We use Root Mean Square Error (RMSE) in kWs and Mean Absolute Error (MAE) in kWs as forecast metrics. During our deployment period, on sunny days, the RMSE is 0.25 kW and the MAE is 0.22 kW on average. On cloudy days, the RMSE is 0.37 kW and the MAE is 0.32 kW on average. cloudy days occur more frequently than sunny days, we use a weighted average to estimate the overall RMSE and MAE. Overall, the RMSE is 0.33kW and the MAE is 0.29 kW. We optimize detection performance under varying conditions using Pareto front optimization to balance power and bandwidth constraints. Considering the impact of volatile weather on energy and connectivity, we incorporate energy forecasts and real-time bandwidth monitoring to ensure efficient operation even in adverse conditions. Figure~\ref{fig:pareto} shows the Pareto front from our YK river and KN river datasets, noting that the z-axis represents the normalized detection performance metric ranging from 0 to 1. This is bounded by power consumption and satellite network connectivity. The results demonstrate the optimal trade-offs between power, bandwidth, and performance in real-world experiments.

\section{Further Discussion}
\label{sec:discussion}
\textbf{Scalability of SALINA.} Scalability is a critical aspect of the SALINA system, ensuring its deployment across various remote wild ecosystems with minimal adjustments. The modular design of SALINA allows for seamless integration of additional sonar units and edge devices. By leveraging cloud-based resources, the system can dynamically allocate computational power and storage based on real-time demand~\cite{xu2018enhancing, gunasekaran2021cocktail, zeng2020distream}, thereby supporting larger deployments without compromising performance. Furthermore, the use of containerized microservices enables easy replication and distribution of processing tasks across multiple machines, ensuring load balancing and fault tolerance~\cite{yu2023following}. This scalability is essential for long-term monitoring and large-scale studies.


\textbf{Integration with large language models and vision language models.} Incorporating large language models (LLMs) and vision language models (VLMs) enhances SALINA's analytical capabilities, enabling sophisticated data interpretation and interaction. LLMs, such as GPT-4o~\cite{openai2024gpt4o}, Llama 3~\cite{touvron2023llama}, can assist in generating comprehensive reports, translating raw sonar data into actionable insights, and facilitating seamless communication between technical and non-technical stakeholders. Meanwhile, VLMs~\cite{radford2021learning, liu2023llava, liu2023pllava}, which combine visual and linguistic data processing, can improve the accuracy and contextual understanding of sonar imagery. These models can annotate sonar data with rich descriptive metadata, improving the semantic understanding of detected objects and events. This integration is promising for advanced applications such as automated anomaly detection, behavioral analysis, and interactive query-based data exploration, further extending the utility of the SALINA system.

\textbf{Implications on related applications.} SALINA can be adapted for diverse remote monitoring scenarios, such as tracking endangered species, assessing water quality, or supporting flood early-warning systems. For instance, deploying SALINA in flood-prone river basins could enable continuous monitoring of water levels, flow rates, and underwater sediment movements. By integrating with local hydrological sensors, SALINA can provide real-time data to detect rising water levels or sudden changes in flow patterns, offering an early indication of potential flooding events. This real-time feedback would allow authorities to issue timely alerts, evacuate at-risk areas, and implement mitigation strategies.

\section{Conclusion}
\label{sec:conclusion}
In this paper, we presented SALINA, a sustainable live sonar analytics system designed for deployment in wild ecosystems. Our system addresses key challenges such as the absence of relevant datasets and models, the need for real-time processing under resource constraints, and the integration of sustainable energy sources. SALINA incorporates advanced data preprocessing, tailored DNN models for both on-premise and cloud inference, and optimizing network performance and power use.
Through extensive real-world deployments and evaluations, our results indicate that SALINA can effectively support continuous underwater surveillance and provide valuable insights for wildlife monitoring and resource management. The methodologies and findings reported in this study offer a comprehensive framework for deploying similar acoustic data analytics systems in various challenging environments.


\begin{acks}
This research is supported by an NSERC Discovery Grant, a British Columbia Salmon Recovery and Innovation Fund (BCSRIF\_2022\_401), and a MITACS Accelerate Cluster Grant. Jiangchuan Liu is the corresponding author. The authors would like to thank Meredith Adams and Ravi Camire for their assistance in installing and configuring the system. We are also grateful to experiment.com and the Haida First Nation for their great support. Finally, we would like to thank the anonymous reviewers and our shepherd for their constructive feedback and guidance.
\end{acks}

\balance
\bibliographystyle{ACM-Reference-Format}
\bibliography{sonar_update}

\end{document}